\documentclass[preprint]{elsarticle}
\usepackage{natbib, geometry, fleqn, graphicx, txfonts, hyperref, endfloat}
\usepackage{algorithmic}
\usepackage{algorithm}
\usepackage{array}
\usepackage{textcomp}
\usepackage{stfloats}
\usepackage{url}
\usepackage{verbatim}
\usepackage{cite}

\journal{Energy Reports}

\begin{document}

\begin{frontmatter}
\title{Smart Grids: A Comprehensive Survey of Challenges, Industry Applications, and Future Trends}

\author[1]{Jadyn Powell}
\ead{jpowel45@uwo.ca}

\author[2]{Alex McCafferty-Leroux\corref{cor1}%
}
\ead{mccaffea@mcmaster.ca}

\author[2]{Walid Hilal}
\ead{hilalw@mcmaster.ca}

\author[2]{S. A. Gadsden\corref{cor2}%
}
\ead{gadsden@mcmaster.ca}

\cortext[cor1,cor2]{Corresponding author}

\affiliation[1]{organization={Western University},
addressline={1151 Richmond St},
postcode={N6A 3K7},
city={London},
country={Canada}}

\affiliation[2]{organization={McMaster University},
addressline={1280 Main St. W},
postcode={L8S 4L7},
city={Hamilton},
country={Canada}}

\begin{abstract}
With the increased energy demands of the 21\(^{st}\) century, there is a clear need for developing a more sustainable method of energy generation, distribution, and transmission. The popularity of Smart Grid continues to grow as it presents its benefits, including interconnectivity, improved efficiency, the ability to integrate renewable energy sources, and many more. However, it is not without its challenges. This survey aims to provide an introductory background of smart grids, detail some of the main aspects and current challenges, and review the most recent papers and proposed solutions. It will also highlight the current state of implementation of the smart grid by describing various prototypes, as well as various countries’ and continents’ implementation plans and projects. 
\end{abstract}

\end{frontmatter}

\section{Introduction}
The electrical grid, pivotal in producing, transmitting, and distributing electricity, is instrumental to economic and social development. Its central role lies in spatially allocating electricity \citep{grid2030, roadmap, energyuk, enabler}. Despite being lauded as one of the paramount engineering feats of the 20th century \citep{refd} and heavily relied upon by consumers, current electricity delivery methods are rigid.

Today's delivery systems, composed of various transmission and distribution networks, supply consumers with electricity from centralized generation stations \citep{demandside}. Yet, this expansive, intricate system finds it challenging to meet the escalating demand for real-time, reconfigurable, and adaptive functionalities. Its continuous operation largely hinges on human intervention, owing to the absence of automated self-correcting features crucial in today's dynamic environments \citep{cunjiang}. While there are ongoing incremental improvements in these systems, a thorough revamp of the infrastructure is indispensable to cater to the soaring demand for smart systems.

Enter the smart grid (SG), heralding a paradigm shift in electricity delivery. The SG integrates modern telecommunication and sensing technologies to enhance electricity delivery strategies \citep{blumsack}. Unlike the traditional unidirectional grid, the SG introduces a bidirectional framework, facilitating a bidirectional flow of information and electricity \citep{6099519}. This evolution fosters increased customer engagement, enables the grid to operate more collaboratively, improves monitoring, enhances automation, and ensures widespread access to information \citep{blumsack}.

With the SG's integration of advanced monitoring and sensing technologies, less human intervention is required. The grid evolves to possess self-healing abilities, enhanced demand response, and smoother renewable source integration. These advancements not only offer consumers and providers more flexibility but also open the doors to technologies currently incompatible with the existing grid infrastructure \citep{hledik2009green}.  

In light of the pressing need to combat climate change, the SG offers a promising avenue to slash carbon emissions in the power sector and seamlessly integrate renewable energy. Current power systems may become untenable, especially with energy production demands projected to soar by 70\% by 2040. This spike underscores the urgency to develop an efficient, sustainable power system to meet future challenges \citep{energyandair}. 

The objective of this paper is to furnish a comprehensive overview of the latest research on SGs, offering clear analyses of diverse research trajectories. We aim to collate recent contributions, chart advancements in the domain, and shed light on the SG's potential trajectory. The remainder of the survey unfolds as follows: Section 2 delves into related surveys in the domain. Section 3 elucidates the background of the SG. Section 4 tackles cyber-security issues, threat classifications, and proposed solutions. Section 5 discusses the imperatives and challenges of interoperability, while Section 6 delves into renewable energy integration. Section 7 provides insights on SG industry applications and future trends, and Section 8 concludes the discourse.

\section{Related Publications}

The evolution of SG research has seen numerous contributions spanning various areas. Over time, there has been a broad range of research literature published on SGs, including papers focused on broad-scale information, technologies, and architecture. 

One of the first few survey papers provided about SG was published in 2010 and provides an overview of SG, including its drivers, evolution, standards, and research, development, and demonstration (RD\&D) \citep{hassanf}. Since then, several more survey papers have been published. This 2016 publication from I. Colak \textit{et al.} \citep{ref12} provides a more updated overview of various SG components, such as cybersecurity, renewable energy integration, and interoperability. More recent papers continued to be published, such as the following from 2020 \citep{ref13}. This research provides an in-depth background of the SG’s definition, architecture, functions, and possibilities. In a recent 2022 paper, Judge \textit{et al.} provides a modern overview of the SG's performance and impacts and details the integration of renewable energy sources \citep{ref14}. 

There are also other publications that thoroughly investigate a few of components of the SG. A 2020 paper provided a comprehensive overview of renewable energy integration \citep{ref15}, as well as Kawoosa and Prashar's work from 2021, which provides a comprehensive review of cyber security including classifications, threats, and proposed solutions \citep{ref16}.

Of the existing survey papers, two prominent limitations were present. Either the content was overly restrictive, or overly inclusive. The papers were very detailed about a small component of SG, or they provided a general overview of SG. Typically, these works do not present enough detail about the SG. Another significant issue is that few publications highlight the current state of implementation of SG worldwide. Regardless of various shortcomings, each paper presents valuable information for readers.

In this survey, the work of previous authors is extended, providing a thorough overview of SGs, their purpose, and benefits. We also address the limitations of the previously mentioned surveys, such as the inclusion of updated technologies, detailed but limited content, and emphasis on the overall state of the SG implementations. This paper also serves as an updated, comprehensive review of essential components of the SG, as well as its current state of implementation. This will be achieved by defining the SG, detailing its importance, and exploring its main components. The main components of the SG are also identified and explained, as well as the challenges and most updated proposed solutions. The current state of the SG is also highlighted, presenting some of the main countries' and continents' projects and the progress of implementing the SG. 

\section{Introduction to Smart Grid}

\subsection{Traditional Electrical Grids}

The current electric power grid is a complex, physical infrastructure used for the distribution of electricity \citep{refm}. There are three main systems within the electric grid, including the electrical generation, transmission, and distribution systems. This integrated network is used to deliver electricity to consumers and includes the power plants used to generate electricity, the substations used to transform voltages, and the distribution facilities to deliver electricity from substations to consumers.

The physical infrastructure lacks automation, relying heavily on employee services for maintenance and repairs and continuously proves unsustainable as future technologies continue to develop, such as advanced metering, and remote monitoring. Despite being widely relied on, the current electric grid has a significant number of uncertainties, including aging infrastructure, and poor resiliency to disturbances. It is also nonlinear and incredibly complex \citep{ref17}. 

With the increased demand for more climate sustainable actions, the integration of renewable energy sources into the power grid is a necessary step. Unfortunately, this step will add even more complexity to the already complex system and present more challenges to several controllers at all levels of the power grid \citep{ref17}. This clearly indicates that a more sustainable alternative to the electric grid, or heavy augmentation of it, will be necessary in the near future. 

\subsection{Smart Systems}

Smart systems are used to embed technology into already existing systems and processes to increase their efficiency and automation. They use incorporated functions such as sensing, actuating and controls to analyze situations making predictive and adaptive decisions.

Smart systems are able to learn, reason, control, and perceive themselves and their environment \citep{ref21}. They are able to self-organize and provide communication between various elements throughout the system \citep{ref21}. They can continuously perceive their surrounding environment, allowing the systems to update their internal knowledge, which is used for effective decision making \citep{ref21}. Through reasoning capabilities, smart systems are adaptable to both new states and objectives of their environment \citep{refe}.

Smart systems have various applications and advantages, including improved security, interconnectivity, and improved functionality. The integration of smart systems to the current electrical grid will allow for solutions to various rising issues and for continued modernization in the future.

\begin{figure}[ht]
\centering
\includegraphics[width=8.5cm, height=6cm]{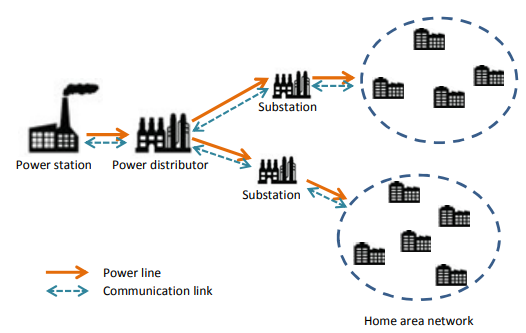}
\caption{Basic SG Structure \citep{refua}}
\label{fig:sg}
\end{figure}

\subsection{Smart Grid}

The SG is a new, modern grid combining smart systems with the current electrical grid. It uses a bidirectional flow of both information and electricity, creating an automated delivery mode \citep{reff}. It aims at improving energy efficiency and demand-side management, and reinforcing reliable grid protection through self-healing methods \citep{6099519}. It combines multiple bidirectional smart devices (e.g. sensors, actuators, and meters) which will provide real-time monitoring, balancing, and control at all times with high accuracy \citep{ref20,ref21}.

While the traditional grid can only distribute electrical power, the SG is able to make, communicate, and store its decisions, augmenting the current grid and enabling cooperative, responsive, and organic functions \citep{refh}. It incorporates many unique technological features and increased monitoring, which makes the behavior of both the consumers and suppliers more apparent and therefore easier to understand \citep{ref23}. Figure \ref{fig:sg} above outlines the basic structure of a SG, showing the bidirectional flow of information and sequential power distribution \citep{refua}.

\subsubsection{Control and Monitoring}

The grid is operated, monitored, and controlled using Information and Communications Technologies (ICT). ICT enables energy companies to control the power demand, allowing for reliable and efficient power delivery at a reduced cost \citep{refg}. Based on information received from consumers, the SG prepares and executes streamlined operations using bidirectional communications between electric power companies and consumers \citep{ref24}. We can consider SG as being intelligent as it applies protection systems of the central/grid control, grid computing, complete diagnostic monitoring of transmission equipment, and self-healing power system networks using distributed computer agent \citep{ref22}. By using a Supervisory Control and Data Acquisition (SCADA) system, these technologies can be facilitated \citep{ref22}. SCADA is a specialized control system used to remotely monitor, control, and manage critical processes and equipment. SCADA systems collect data from sensors and instruments in the field, transmit and process the data. They allow operators to make informed decisions, initiate control actions, and respond to alarms and faults through the providing of real-time information about the status of processes and equipment.

\subsubsection{Smart Meters}

Smart meters are another unique component of the SG which enables advanced metering infrastructure (AMI). This allows for more data to be collected and in a much more effective manner. Smart meters collect data every minute, whereas old metering data was recorded hourly or monthly. To improve metered grid data (i.e. voltage and current phasors) accuracy at more frequent sampling intervals, phasor measurement units (PMUs) can be applied. The PMU's implementation into SG has been extensively researched (see \citep{refi}), significantly contributing to the SG's advancements in the control, estimation, and security of power grids. As a comparison, modern PMUs collect up to 60 data points per second, whereas the current SCADA systems collect a single data point every 1 to 2 seconds \citep{ref17}. Applying the AMI in the power distribution system and the PMU in the power transmission grid will provide the power grid a much more in-depth look at grid performance compared to the data available from SCADA technology \citep{ref17}. Incorporating the use of additional smart technology in consumer’s homes will expand control and monitoring of multiple devices connected to the power grid. 

\subsubsection{Anticipated Benefits}

There are many anticipated benefits from the integration of SGs, originally listed by the National Institute of Standards and Technology (NIST) \citep{ref94}, and subsequently redefined by Fang \textit{et al.} \citep{6099519}. The benefits are categorized and described below.

\begin{itemize}
    \item Grid Reliability and Quality Improvement
        \begin{enumerate}
            \item Improving the quality and reliability of power transmission
            \item Improving resilience to grid related disruptions (e.g. power and data)
            \item Presenting opportunities for improving grid security
        \end{enumerate}
    \item Grid Efficiency and Optimization
        \begin{enumerate}
            \item Optimizing the utilization of facilities, thereby avoiding back-up power plant construction
            \item Enhancing the capacity and efficiency of existing power networks
            \item Automating maintenance and various operations
        \end{enumerate}
    \item Grid Modernization and Adaptation
        \begin{enumerate}
            \item Enabling self-healing reactions and predictive maintenance when subjected to disturbances
            \item Advancing the renewable energy source deployment
            \item Distributed power source accommodation
            \item Standardizing new energy storage alternatives and the transition to plug-in electric vehicles
        \end{enumerate}
    \item Environmental Sustainability and Efficiency
        \begin{enumerate}
            \item Reducing greenhouse gas emissions through electric vehicle normalization and utilizing modern power sources
            \item Minimizing inefficient generation during periods of peak-usage, and therefore oil consumption 
        \end{enumerate}
    \item Market and Consumer-Related Enhancements
        \begin{enumerate}
            \item Increasing consumer choice 
            \item Enabling the introduction of new markets, services, and products
        \end{enumerate}
\end{itemize} 

The benefits as described by NIST in \citep{ref94} offer numerous benefits to the individuals, countries, and companies that adopt SGs. With environmental, security, and operational benefits augmented into the power grid, consumers can also expect to see an increased amount of choices offered to consumers considering their energy consumption, as well as new markets, services, and products. An example that applies to the United States (US) and many other countries, the implementation of SGs will be important for reducing the dependence on foreign oil and increasing clean energy production \citep{ref94}, therefore creating jobs and new commodities, manufacturing methods, etc. from that research. The first four major points on the list above are directly related to the following three sections of this survey, being cybersecurity, interoperability, and renewable energy integration in SGs. Focusing research efforts on implementing these aspects into the power grid will continue to result in achieving the outlined benefits, vital to a sustainable living and future. It was remarked by the Department of Energy (DOE) in 2008 \citep{refn} that if modern power grids were 5\% more efficient than they currently are, the energy savings would be equal to the effect of eliminating the emissions of 53 million automobiles.

\section{Cyber Security}

As electrical grids become more sophisticated, the number of risks they are vulnerable to increases \citep{ref28}. Some risks and threats are linked to the incorporation of digital communications, including cybersecurity and data privacy, reliability, and technical failures, while others come from changes in how power companies and their customers interact \citep{ref17}.

The SG allows for close interaction at all levels of the power distribution, consumption, generation, and transmission systems. This means there is a high degree of coordination and communication amongst the various levels of the power grid system. This close interaction presents many more opportunities and the increased possibility of cyberattacks and cascade failures, which could affect all of the aforementioned systems \citep{ref17}. This could lead to potentially catastrophic consequences, including energy market chaos, SG IT infrastructure failures, power blackouts, dangers to human safety, and damaged customer devices \citep{ref17}. Less severe but more frequent consequences are also a possibility, such as small-scale outages \citep{ref17}. The electric grid is essential, so it is necessary to devise and implement defense systems capable of supervising mass amounts of data, evaluating system status, identifying failures, predicting threats, and suggesting corrections \citep{ref17}.

SG is considered by Gunduz and Das \citep{ref30} to be one of the most consequential applications of the Internet of Things (IoT). An IoT network enables devices the ability to communicate either directly, or through an internet gateway \citep{ref29,ref30}. IoT-based SG systems are large, critical infrastructures, with complex architectures which are vulnerable to cyberattacks. Utilizing IoT applications is convenient for SGs, however, it poses several vulnerabilities. Considering that the monitoring and control operations are performed on internet-based protocols, the SG could potentially be appealing as critical infrastructure to malicious entities \citep{ref30}. Due to this appeal, it is necessary to examine weaknesses in an SG component's security systems, as well as potential cyber security threats within its infrastructure \citep{ref30,ref31}.

\subsection{Classifying Types of Attacks}

\begin{figure}[ht]
\centering
\includegraphics[width=8.5cm, height=6cm]{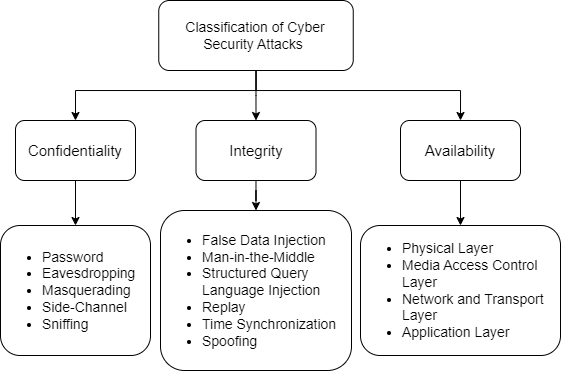}
\caption{SG Classification of Cyber Security Attacks}
\label{fig:chart}
\end{figure}

Due to the enormity of the system, both in terms of geographical distribution and complex infrastructure, identifying and classifying each individual possible attack would be nearly impossible. Instead, three principal security objectives are defined and must be incorporated into SG. These three security objectives are confidentiality, integrity, and availability, also known as the CIA triad, as seen in Figure \ref{fig:chart}. The CIA triad is essential for both the management, operation, and protection of the system and communication infrastructures \citep{refj}.

Cyberattacks in SGs target no less than one of the CIA triad objectives and are conducted to exploit information to use to attackers’ advantage or to harm others \citep{ref30}. Some attacks are coordinated to exploit various components of the SG and to launch simultaneous attacks. The most challenging attack to defend against, a coordinated attack can surpass common defences, requiring multilayer security solutions with robust approaches \citep{ref30}, such as those analyzed in \citep{refa,refb,refc}. These types of attacks target all the components, requirements, and security objectives in an SG \citep{ref30}.

Cyberattacks such as these can occur at numerous communication layers within SG \citep{ref38}. These layers, the network/transport layer, the MAC layer, and the physical layer are all affected differently by various types of attacks \citep{ref30}. Identifying the types of cyber security threats enables effective and appropriate countermeasures \citep{ref32}. Security procedures achieved by investigating the cyber-security requirements according to attack types and the communication layers affected will produce effective solutions for the security of the SG \citep{ref30}.

\subsection{Confidentiality Attacks}

Confidentiality attacks do not intend to change transmitted information over power networks. Instead, these attacks are used with the intention of achieving access to information by unauthorized parties \citep{ref33}. This information includes the customer’s account details and power usage. If malevolent entities were to obtain customer account information, they may be able to commit identity theft, privacy invasions, phishing, and selling customer information. If they obtain power usage information, they may be able to commit billing fraud and energy theft.

Confidentiality attacks are typically considered to have minimal effects on the functionality of the communications in SG. However, the importance of customer privacy has gained more attention as more progress is made with the SG \citep{ref30}. Table 1 below describes various attacks described in this paper that target confidentiality, such as password attacks, masquerade attacks, and side-channel attacks.

\begin{table}[ht]
\caption {Types and Descriptions of Confidentiality Attacks}
\centering
\resizebox{\columnwidth}{!}{%
\begin{tabular}{l|l}
Type of Attack        & Description of Attack                                                                                                                                                   \\ \hline
Password Attacks      & \begin{tabular}[c]{@{}l@{}}Attempts to gain access by \\ using an authorized person's password\end{tabular}                                                             \\ 
\\
Eavesdropping Attacks & \begin{tabular}[c]{@{}l@{}}Intercept wireless transmission \\ on LAN, sniff IP packets in SG \\ networks, or eavesdrop on messages \\ shared between nodes\end{tabular} \\ 
\\
Masquerade Attacks    & \begin{tabular}[c]{@{}l@{}}Attackers pretend to have authorization \\ in order to access privileges\end{tabular}                                                        \\
\\
Side Channel Attacks  & \begin{tabular}[c]{@{}l@{}}Performed to obtain cryptographic \\ keys\end{tabular}                                                                                       \\ 
\\
Sniffing Attacks      & \begin{tabular}[c]{@{}l@{}}Performed to gain access to TCP/IP \\ packets or PMU contents \\ transmitted over a network\end{tabular} \\ 
\\                        
\end{tabular}%
}
\end{table}

Smart meters measure and store large quantities of data and autonomously transport it to providers, consumers, and utility companies. If attackers intercept the private information of the consumer, it can be abused in many ways \citep{ref30}. For example, it can be used to learn what appliances they use, keep track of their lifestyle, and by studying their power usage, determine whether they are at home or not \citep{ref30,ref37}. It may even be possible to differentiate between various activities, such as watching television or sleeping \citep{ref17}. Power usage analysis of businesses may also be able to indicate attackers’ changes in business operations. Confidentiality is a leading issue for business owners and users \citep{ref33}.

Password attacks have various methods, such as password guessing or sniffing, social engineering, and dictionary attacks \citep{ref30}. Social engineering is an especially popular method as it is used to penetrate systems through utilizing social skills, as opposed to technical attacks \citep{ref30,ref35}. 

Eavesdropping attacks are a passive attack type, which damages data confidentiality \citep{ref30}. Eavesdropping attacks intercept wireless transmissions on local area networks (LAN), messages shared between communication network nodes, or sniff IP packets in SG networks \citep{ref30}. 

When masquerading attacks occur, attackers pretend to have authorization to access privileges \citep{ref30}. Common applications of these types of attacks are identity spoofing and impersonation \citep{ref36}. Identity spoofing attacks enable attackers to imitate an authorized individual, without the possession of a user's password \citep{ref35,ref37}. Such attacks include Man-in-the-Middle (MITM), message replay, and network spoofing. Spoofing attacks involve modifications to the parameters of SG devices and can include procedures of Media Access Control (MAC), IP, and address resolution protocol (ARP) spoofing \citep{ref30}. 

Side-channel attacks are performed to obtain cryptographic keys \citep{ref30}. Common types of this attack include timing attacks, power analysis attacks, and electromagnetic analysis attacks \citep{ref37,ref35}. Devices such as smart meters and home appliances are especially vulnerable to these attacks and can result in violations of privacy, information about usage, and administrative access \citep{ref30}.

Using various tools for packet analysis or sniffing, an attacker can obtain TCP/IP packets of smart meters sent over the network or PMU contents \citep{ref30}. TCP/IP packets, PMU, and smart meters are the primary targets of sniffing attacks, and lacking encryption, important information can be detected or collected by attackers \citep{ref30}. The current practice for smart meters is the implementation of an X.509 certificate, which is a standard for device identification and cryptographic session establishment over the internet \citep{ref17}. However, the cryptographic keys are static for each device, meaning that new methods should include a key management solution that can update cryptographic keys periodically \citep{ref17}. 

\subsection{Integrity Attacks}

Attacks that target integrity aim to modify and/or disrupt the exchange of data within SG \citep{ref38}, which can lead to safety issues involving equipment or people \citep{ref39}. Integrity attacks focus on the content of the originally sent data (e.g. billing data, account data, control commands, sensor and voltage values, operating status of devices, etc.), and once intercepted are modified, reordered, or delayed \citep{ref30,ref40,ref41}. Data integrity checks are often utilized to mitigate these attacks since this information is both personal to customers and valuable to companies. There are also fault-tolerant methods against data integrity attacks, as discussed in \citep{refk,refl}. Though they are generally considered to be not as "brute-force" as other methods, attacks targeting data integrity are more sophisticated than attacks, targeting other categories of the CIA triad \citep{ref38}. Table 2 below describes various attacks described in this paper that target integrity.

\begin{table}[ht]
\caption{Types and Descriptions of Integrity Attacks}
\centering
\resizebox{\columnwidth}{!}{%
\begin{tabular}{l|l}
Type of Attack               & Description of Attack                                                                                                                                                                                                                                                            \\ \hline
FDI Attacks                  & \begin{tabular}[c]{@{}l@{}}Incorrect data is introduced into NAN \\ measurements or smart meters.\\ They damage the integrity of \\ monitoring and measurement subsystems, \\ resulting in cascaded poor judgement \\ in the SG network\end{tabular}     \\
\\
MITM Attacks                 & \begin{tabular}[c]{@{}l@{}}Performed to damage the transmission \\ between the smart meter and \\ data concentrator units\end{tabular}                                                                                                                                   \\
\\
SQL Injection Attacks        & \begin{tabular}[c]{@{}l@{}}Transforms databases through the injection \\ of script commands or malicious queries \\ into the database, resulting in command over \\ the system, modified or erased data, and \\ additional manipulated data\end{tabular}                                  \\
\\
Replay/Playback Attacks      & \begin{tabular}[c]{@{}l@{}}Retransmits or delays messages \\ after acquiring them through masquerading \\ attacks. Initiated so attackers can direct \\ energy to different locations and cause \\ physical damage to the system\end{tabular}                              \\
\\
\begin{tabular}[c]{@{}l@{}}Time Synchronization\\ Attacks\end{tabular}           & \begin{tabular}[c]{@{}l@{}}Targets timing data in SG, mainly PMU\\ and WAPMC. TSA may generate\\ incorrect location errors, triggering a\\ false alarm indicating an issue, which\\ can cause communication line interruptions,\\ resulting in cascading faults\end{tabular} \\
\\
Spoofing Attacks             & \begin{tabular}[c]{@{}l@{}}An attacker can impersonate other devices \\ by exploiting the openness of the \\ address fields on the MAC layer, enabling \\ the ability to send false information \end{tabular}                                                                               
\end{tabular}%
}
\end{table}

The false data injection (FDI) attack is emerging as one of the most dangerous types of cyber-attacks for SG. FDI occurs when bad data is injected into neighborhood area network (NAN) measurements or smart meters, targeting SG infrastructure \citep{ref30,ref43}. The objective of the FDI is to damage the integrity of both monitoring and measurement sub-systems, resulting in cascaded poor judgment throughout the SG network \citep{ref44}. Monitoring measurement manipulation would result in incorrect evaluations of the operating state of the system and therefore the destabilization of the SG, and to imprecise operational actions, such as pricing, planning, self-healing, or general flexibility \citep{ref30}. Oozeer and Haykin comment that conventional malicious data detection and state estimation techniques have been applied to detect bad data in energy system state estimators and reduce observation errors \citep{ref26}. They are unable to, however, detect FDI attacks \citep{ref26}, leading to the development of more advanced or intelligent methods. If it is assumed that an attacker has already compromised one or more meters, Y. Liu \textit{et al.} \citep{ref45} suggested that the attacker is able to inject counterfeit data into the SCADA center, simultaneously passing the data integrity check utilized in the current state estimation process \citep{ref38}. The load redistribution attack \citep{ref46} is a variation of FDI, where only line power flow and load bus injection measurements and are influenced in the attack \citep{ref38}. Research in \citep{ref46} demonstrates that these attacks can be considered as realistic FDI attacks, having constrained access to specific meters \citep{ref38}. 

Data integrity attacks exploit vulnerabilities to corrupt processes in SG \citep{ref30,ref66}. Common methods of such attacks include MITM attacks and Structured Query Language (SQL) injections, each having their own variations. For MITM, these include compromised certificates, modified packet source/destinations, and route table poisoning \citep{ref30,ref66}. In SG applications, the data concentrator unit is connected to home area network (HAN) smart meters. Utilizing data modification attack methods (typically MITM), attackers are able to damage the data transmission between these two sub-systems, negatively influencing the accountability of the system and the CIA triad \citep{ref30}.

SQL injection attacks are intended to inject script commands and alter databases\citep{ref30}.
Smart meters continuously forward power consumption data to secure databases for both utilities and users, a prime target for these types of attacks. Malicious queries are injected into the database in order to add manipulated data, modify or delete existing data, or gain authority over the system \citep{ref30}. These injections can result in disruptions to SG functions and even eventually result in blackout. Unless the queries formed by the database users are sufficiently validated before insertion, SQL injection can occur \citep{ref30}.

Replay or playback attacks are designed to direct energy to a separate location in the grid and cause physical harm to the system \citep{ref30,ref33}. Similar to previously discussed attacks, there are a variety of types of replay attacks, including covert attacks, which are their closed-loop versions \citep{ref47} existing in short lengths of time and having a particular frequency \citep{refo}. Replay attacks are likely to cause extremely serious effects on system stability, on global or localized scales across the SG, with losses being accumulated from either the delay itself or the attack signal. These attacks can be counteracted on authentication or fault detection/estimation levels. Numerous efforts have been proposed to counteract these attacks and preserve stability, such as in Abdelwahab \textit{et al.}'s active detection using watermarking \citep{refp} and Pavithra and Rekha's fault detection and fast encryption algorithm \citep{refq}. Replay attacks occur when an entity acting as the primary source obtains the network traffic, and forwards the data to a destination. Following the acquisition of data through masquerading attacks, replay attacks are intended to to delay or re-transmit messages \citep{ref30}. Data can be injected into the system by attackers without significant modifications to measurable outputs, as well as target unencrypted sensors to initiate the replay attacks. Monitoring sensor outputs, attackers repeat them while injecting their attack signal \citep{ref30}. Not only are false control signals are injected into the network in replay attacks, but attackers are able to analyze and access the data that is transmitted between devices and meters to obtain the target’s energy generation and usage characteristics \citep{ref30}.

Time synchronization attacks (TSAs) primarily target timing data in SG applications \citep{ref30}. Event location, fault detection in transmission lines, monitoring voltage stability, and other applications of the PMU or WAPMC can be influenced by TSA, as these processes are heavily reliant on exact timing \citep{ref37}. For instance, global positioning system (GPS) spoofing, a type of TSA, aims at imitating a GPS signal, providing false times or locations that are typically used for breaches in defense \citep{ref48}. TSAs may generate incorrect location errors and trigger false alarms, indicating an issue, as proven by the outcome demonstrated in \citep{ref49}. Gunduz and Das suggest that this false alarm can cause interruptions in communication lines, possibly triggering cascading faults in a SG \citep{ref30}. 

In the SG, spoofing is particularly harmful as it targets two categories from the CIA triad: availability and integrity. Exploiting the openness of address fields in a frame on the MAC layer of the SG, an attacker utilizing spoofing can disguise themselves as other devices to send false information from \citep{ref38}. Premaratne \textit{et al.} explain in \citep{ref50} that in power substation networks, malicious nodes are able to broadcast forged ARP packets to collapse connections of all intelligent electronic devices (IEDs) to the substation gateway node.

\subsection{Availability Attacks}

Attacks targeting availability, referred to as denial-of-service (DoS) attacks, aim to corrupt, block or delay communications in SGs \citep{ref38,ref51}. A DoS attack can impair the operation of electronic devices and severely degrade a power system's communication performance, since it's expected to be consistently accessible \citep{ref38}. Attackers flood transmission lines in the network with large volumes of traffic, which results in the loss of legitimate data packets and thus not processed \citep{ref26}. The SG inherits the vulnerabilities in the TCP/IP stack and becomes vulnerable to DoS attacks since it applies TCP/IP stack and IP protocols \citep{ref18}. Since availability is the dominant security requirement for SG, advanced and effective countermeasures must be adopted to defend against these types of attacks \citep{ref30}. Table 3 below describes various attacks described in this paper which target availability.

\begin{table}[ht]
\centering
\caption{Types and Descriptions of Availability Attacks}
\label{tab:my-table}
\resizebox{\columnwidth}{!}{%
\begin{tabular}{l|l}
Type of Attack                                                                   & Description of Attack                                                                                                                                                                                                   \\ \hline
DOS Attacks                                                                      & \begin{tabular}[c]{@{}l@{}}Performed in an effort to corrupt, block, or \\ delay SG communications\end{tabular}                                                                                                 \\
\\
\begin{tabular}[c]{@{}l@{}}Physical Layer \\ Attacks\end{tabular}                & \begin{tabular}[c]{@{}l@{}}Intended to crowd wireless communication \\ lines with noise, blocking connection \\ between users and smart meters\end{tabular}                                                       \\
\\
\begin{tabular}[c]{@{}l@{}}MAC Layer\\ Attacks\end{tabular}                      & \begin{tabular}[c]{@{}l@{}}Performed in an attempt to modify the MAC \\ parameters of a device in order to cause \\ performance degradation of other devices \\ sharing the same communication channel\end{tabular}     \\
\\
\begin{tabular}[c]{@{}l@{}}Network and \\ Transport Layer\\ Attacks\end{tabular} & \begin{tabular}[c]{@{}l@{}}These attacks are launched in order to \\ initiate degradation in end-to-end \\ communication performance\end{tabular}                                                                                   \\
\\
\begin{tabular}[c]{@{}l@{}}Application Layer\\ Attacks\end{tabular}              & \begin{tabular}[c]{@{}l@{}}Primarily focused on the transmission \\ bandwidth in routers, computers, or \\ communication channels, intended to exhaust \\ computer resources, such as its CPU or \\ I/O bandwidth\end{tabular}
\end{tabular}%
}
\end{table}

DoS attacks can occur at various of communication layers within the SG, all having varying affects and severities \citep{ref38}, as described below.

\subsubsection{Physical Layer Attacks}

An effective method in launching physical layer attacks, especially for wireless communications, is with channel jamming, extensively covered in \citep{ref52,ref53,ref54}. It is quite simple to perform DoS attacks at this layer since it connection with communication channels is required, rather than authenticated networks \citep{ref38}. Wireless jamming is the primary attack in local area systems, since wireless technology will be widely used throughout these networks \citep{ref55,ref56,ref57,ref58,ref59}. These attacks aim to congest wireless communication lines with noise-corrupted signals, blocking connection between smart meters and users \citep{ref37}. Blocking data packages from being received, and communication channels being continuously seen as busy by routers result in a significant drop in smart meter performance and reliability \citep{ref30}. Research performed by Z. Lu \textit{et al.} \citep{ref60} reinforce this, showing that jamming attacks can result in a variety of impairments to power substation systems, ranging from the delayed delivery of time-critical messages to total denial-of-service. Stravrou and Keromytis \citep{refr} detail that an effective approach to these attacks is dispatching random unauthenticated packets to all wireless stations in the network, preventing attacks from following. With their novel spread spectrum approach, they effectively reject DoS attacks of differing degrees of sophistication and size, without the requirement of software or hardware augmentation \citep{refr}. Another method proposed by Lee and Gerla \citep{refs} offers security on the physical layer from random frequency hopping, similar to the spread spectrum approach. 

\subsubsection{Medium Access Control (MAC) Layer Attacks}

The MAC layer is responsible for point-to-point communication \citep{ref38}. A compromised device could be used by a malicious entity to deliberately modify the MAC parameters, resulting in performance degradation of multiple devices that are coupled to a single communication channel. Therefore, MAC layer complications often contribute a relatively weaker class of a DoS attack \citep{ref38}.

\subsubsection{Network and Transport Layer Attacks}

Considering the TCP/IP protocol model, the transport and network layers must provide reliability control for information delivery over multi-hop communication networks \citep{ref38}. At these layers, the DoS attacks can result in deteriorated performance of end-to-end communication \citep{ref38}, including worm propagation attacks over the Internet and distributed traffic flooding \citep{ref61,ref62,ref63}. Several studies have been conducted to asses the impact of transport- and network-layer DoS attacks on the power system performance \citep{ref38}. D. Jin \textit{et al.}, \citep{ref64} investigated the influence of a buffer-flooding attack on a DNP3-based SCADA network with real software and hardware, demonstrating the vulnerability of SCADA systems to DoS attacks \citep{ref38}. 

\subsubsection{Application Layer Attacks}

Lower-layer attacks, which occur on the application layer, mainly focus on transmission bandwidth in routers, computers, or communication channels, and are intended to exhaust computer resources, like CPU or I/O bandwidth \citep{ref38}. S. Ranjan \textit{et al.} \citep{ref65} demonstrate the overwhelming ability of these attacks towards a computationally limited computer, inducing failure from a continuous stream of computationally heavy demands. The SG is particularly vulnerable to the DoS attack, since its is comprised of innumerable computing and communication modules, some equipped with limited computational abilities \citep{ref38}.

\subsection{Attack Defenses}

While there are standard protocols to help defend the SG against more thoroughly researched and understood confidentiality, integrity, and availability attacks, there has been increased research into the newer attacks on the SG, such as FDI attacks, and further, sophisticated methods of defense. Table 4, seen below, summarizes the papers examined in this section.

\begin{table}[ht]
\centering
\caption{Summary of Attack Defense Papers}
\label{tab:my-table2}
\resizebox{\columnwidth}{!}{%
\begin{tabular}{l|l|l|l}
Year & Reference  & \begin{tabular}[c]{@{}l@{}}Attack \\ Classification\end{tabular} & Method  \\ \hline
\rule{0pt}{4ex}  
2013 & \citep{ref37} & Confidentiality  & Data encryption \\
\rule{0pt}{4ex} 
2015 & \citep{ref36} & Confidentiality  & Authentication process \\
\rule{0pt}{6ex}  
2018 & \citep{ref66} & Confidentiality  & \begin{tabular}[c]{@{}l@{}}Constructing discrete \\ infrastructure for\\ communication\end{tabular}  \\                                                  
\rule{0pt}{4ex}  
2016 & \citep{ref47} & Integrity, Replay & \begin{tabular}[c]{@{}l@{}}Time-stamps and \\ sequence numbers\end{tabular} \\
\rule{0pt}{3ex}  
2015 & \citep{ref67} & Integrity  & Cryptography \\ 
\rule{0pt}{10ex}
2015 & \citep{ref33} & Integrity & \begin{tabular}[c]{@{}l@{}}Increased secure \\ monitoring with PMU,\\ volt-VAR control scheme, \\ power fingerprinting, and \\ reliable network-based \\ methods\end{tabular} \\
\rule{0pt}{12ex}  
2018 & \citep{ref66} & \begin{tabular}[c]{@{}l@{}}Integrity, SQL\\ Injections\end{tabular} & \begin{tabular}[c]{@{}l@{}} Static code checking, \\ positive pattern matching,\\ input type checking, \\ limiting database admission to \\ remote users, implementing \\ penetration tests, avoiding \\ dynamic SQL, and \\ filtering semicolons\end{tabular} \\
\rule{0pt}{4ex}  
2018 & \citep{ref68} & Integrity, MITM & \begin{tabular}[c]{@{}l@{}}Authenticating source and\\ target of information\end{tabular} \\
\rule{0pt}{4ex}  
2010 & \citep{ref69} & Integrity & \begin{tabular}[c]{@{}l@{}}Including PKI and trusted\\ computing methods\end{tabular} \\
\rule{0pt}{4ex}  
2017 & \citep{ref70} & Integrity, FDI & \begin{tabular}[c]{@{}l@{}}Real-time detection using\\ deep learning-based methods\end{tabular} \\
\rule{0pt}{4ex}  
2019 & \citep{ref26} & Integrity, FDI & Entropic state \\
\rule{0pt}{3ex}  
2019 & \citep{ref34} & Integrity, FDI & Entropic state \\
\rule{0pt}{8ex} 
2021 & \citep{ref71} & Integrity, FDI & \begin{tabular}[c]{@{}l@{}}Statistical FDI attack \\ detection approach based on \\ a novel dimensionality \\ reduction and\\ a Gaussian mixture model\end{tabular} \\
\rule{0pt}{6ex}  
2020 & \citep{ref30} & \begin{tabular}[c]{@{}l@{}}Availability,\\ Channel Jamming\end{tabular} & \begin{tabular}[c]{@{}l@{}}Sending random \\ unauthenticated packets to all\\ wireless stations in the network\end{tabular} \\
\rule{0pt}{6ex}  
2018 & \citep{ref66} & Availability & \begin{tabular}[c]{@{}l@{}}Traffic filtering, applying\\ air-gapped network, and\\ anomaly detection\end{tabular} 
\end{tabular}%
}
\end{table}

\subsubsection{Standard Integrity Attack Defenses}

To initiate integrity or confidentiality attacks, the entity performing the attack must have established access to a user's sensitive information and communication networks \citep{ref30,ref38}. To terminate the various types of integrity attacks, end-to-end encryption and authentication schemes are necessary \citep{ref30}. Sanjab \textit{et al.} \citep{ref47} discuss implementing sequence numbers and timestamps as an effectie solution against replay attacks in SG applications. Shapsough \textit{et al.} \citep{ref67} outlines cryptography algorithms as methods to prevent data integrity attacks. Rawat and Bajracharya \citep{ref33} also lists increased security in SG monitoring through utilizing the PMU more frequently, volt-VAR control based schemes, power fingerprinting techniques, and the prevention of integrity attacks towards data by using trusted network connection-based approaches.

Applying measures as discussed by Bedi \textit{et al.} in \citep{ref66}, SQL injection attacks can be significantly lessened across SG networks. They list various SQL defenses such as initiating penetration tests, static code checking, positive pattern matching, input type checking, avoiding the use of dynamic SQL, limiting database access to remote users, and filtering semicolons during type checking \citep{ref66}.

As a means of countering MITM attacks, security gateways can be utilized to encrypt network traffic \citep{ref30}. Security gateways are able to create VPN tunnels for network connection, data encryption at the source, and decryption at the target \citep{ref30}. Typically, the encryption processes occur within hardware solutions. Also, both the target and source of information should be authenticated to interrupt and stop MITM attacks, as shown in \citep{ref68}. Metke and Ekl \citep{ref69} also discussed including trusted computing methodologies and public key infrastructure (PKI), which tethers a user's identity to a public key using a digital certificate. They believed that utilizing PKI technologies is the most desirable solution for SG security, including the "policies and procedures which describe the set up, management, updating, and revocation of certificates", beyond software and hardware improvement \citep{ref69}.

\subsubsection{Standard Confidentiality Attack Defenses}

Processes such as data encryption \citep{ref37}, authentication processes \citep{ref36}, saturating communication channel bandwidth, detecting saturated channels, and constructing discrete infrastructure for the communication of power grid devices \citep{ref30} help to defend against confidentiality attacks such as eavesdropping, masquerade, spoofing, and side-channel attacks, as discussed by Bedi \textit{et al.} \citep{ref66}. 

Countering confidentiality attacks with data encryption and authentication processes are among the more widely applied methods, used for nearly every interaction made on the internet. Concerning smart grids, Gao \textit{et al.} explore a hybrid method of data encryption that compresses and encrypts the data in a single step \citep{reft}. This approach, known as EncryCS, was determined to improve both security and transmission efficiency \citep{reft}. Alternatively, Syed \textit{et al.} pursued homomorphic encryption for secure training of deep learning networks used for fault identification and localization of smart grids \citep{refu}. For authentication, Saxena \textit{et al.} discuss a joint authentication/authorization scheme for confidentiality attacks that requires these processes to be verified simultaneously \citep{refv}, and W. Chim \textit{et al.} propose their tamper-resistant-device-based Privacy-preserving Authentication Scheme for Smart grid networks (known as PASS) \citep{refw}.

\subsubsection{Standard Availability Attack Defenses}

Approaches listed in \citep{ref66} such as big pipes, traffic filtering, applying air-gapped networks, and anomaly detection methods are a few successful solutions to availability attacks. Since DoS attacks are a major threat towards IoT-based SG systems, software solutions on the network layer could also be an effective method in mitigating DoS attacks, as proposed in \citep{ref30}.

The above approaches are both security measures and anomaly detection methods. Availability attack prevention with air-gapped networks involves not having automatic processes or physical connections between systems. Researchers in \citep{refy} utilize an air-gapped method to detect and address such intrusions without controller communication. Another method, traffic filtering, detects and counters availability attacks that analyze the network traffic to detect and filter out false or malicious signals, demonstrated in \citep{refz}. Machine learning based methods can also be applied for availability attack prevention on various layers, such as in research conducted by Jokar and Leung \citep{refx} who proposed a detection and prevention scheme based on ZigBee.

\subsubsection{FDI Attack Defenses}

Extensive research has been conducted to defend against various other SG attacks. However, FDI attacks are one of the newest and most dangerous to the SG, and therefore currently a very popular research topic. 

Y. He, G. Mendis, and J. Wei \citep{ref70} proposed a real-time detection mechanism using deep learning-based mechanisms. At the time of publication, the authors claim that, to the extent of their knowledge, they were the first to propose a real-time FDI attack detection mechanisms. Applying deep learning techniques, they recognize the behavioral patterns of the attacks through a historical analysis of the measurement data, and successfully implement the revealed features for real-time FDI attack detection \citep{ref70}. The results of their simulation illustrate the resilience of their detection scheme to the various amounts of attacked measurements, detection thresholds of the state vector estimator (SVE), and degree of environmental noise levels \citep{ref70}. The results additionally demonstrate the ability of the employed scheme to achieve high detection accuracy in the presence of the occasional operation faults \citep{ref70}.

The entropic state was originally proposed for use in smart grids in \citep{ref26}, and expanded on by the same authors in \citep{ref34}. In a cognitive dynamic system (CDS), the goal of the system is to both control the state and minimize the amount of unknown information in the perceptor. The latter is performed by minimizing this entropic state, which is done by dynamically optimizing the estimation process \citep{ref26}. The authors introduce the entropic states as a "new metric" for the SG, having two key purposes: to be used to detect FDI attacks and to provide an indication of the SG’s health from cycle-to-cycle \citep{ref26}. Oozeer and Haykin demonstrated how the entropic state was able to detect FDI attacks and drive them to a controllable state, under the action of cognitive risk control (CRC). Through task-switching control, the cognitive dynamic system (CDS) was able to enable a new executive with a different set of actions, augmenting the system configuration to bring the risk to a manageable state during an attack \citep{ref26}. The entropic state represents the information gap, quantifying uncertainty and in the case of \citep{ref34}, can be used to detect an attack. The results of simulations showed that their system would prove effective for future SG systems \citep{ref34}. This CRC method has been applied to other systems, such as vehicle-to-vehicle (V2V) communication systems \citep{refaa}, and the CDS method in control systems is a rising research topic.

Shi \textit{et al.} \citep{ref71} proposed an approach aimed at statistical FDI attack detection based on a new dimensionality reduction method and a Gaussian mixture model. They proposed a technique that involves two phases: a dimensionality reduction process in the first phase and a semi-supervised learning process based on the Gaussian mixture model in the second \citep{ref71}. If the Gaussian mixture model outputs exceeds a predetermined threshold, then FDI attacks are identified apart from the usual pattern \citep{ref71}. The simulation results from the tests demonstrate that the proposed scheme detected the FDI attacks with high detection precision and achieved the desired discrimination performance \citep{ref71}.

\section{Interoperability}

\begin{figure}[ht]
\centering
\includegraphics[width=8.5cm, height=7.5cm]{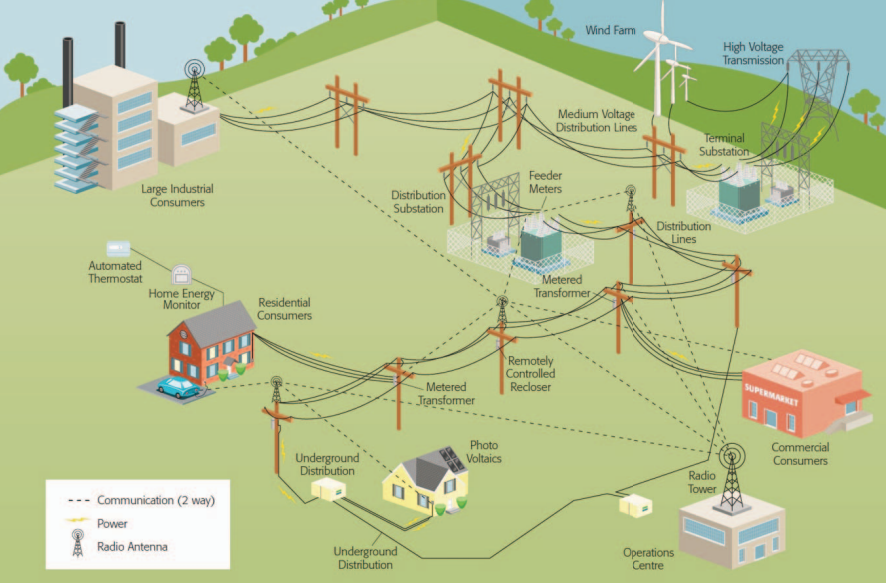}
\caption{Infrastructure of SG interoperability \citep{ref72}}
\label{fig:infra}
\end{figure}

Interoperability refers to the ability to exchange and make use of information. Interoperability presents one of the main features of SG and consists of the capacity of the network technologies, sensors, and electrical devices to use the interchanged information \citep{ref72}. In order for SGs to respond to changes automatically and dynamically in grid conditions, there is a need for sensors to provide real-time status and information \citep{ref20}. This will require seamless connectivity throughout the transmission and distribution system, to enhance the energy flow coordination with real-time analysis and information \citep{ref73}. Interoperability improves the reliability of SG by transferring collected information directly to the equipment, thus improving and protecting the operations of the grid \citep{ref73}. It will improve efficiency and stability of the SG network to avoid overwhelming load on the grid and reduce possible blackouts \citep{ref74}. Figure \ref{fig:infra} \citep{ref72} provides an example of infrastructure for SG interoperability.

The SG’s measurements, communications, and control technologies are used to support system operations maintaining a balance between load demand and electrical generation \citep{ref20}. The geographical disbursement of sensing and measurement devices on the world-scale pose a challenge towards general scalability, availability, and interoperability \citep{ref20}. Interoperability and data exchange between sensors are major challenges for the SG.

An example of a vital operation that heavily relies on the monitoring and measurement of electrical parameters in the distribution and transmission networks is grid control \citep{ref20}. Sensors are implemented to measure the physical parameters of \newline transmission lines, power generation, distribution lines, substations, energy storage, and consumers \citep{ref20}. Such sensors include current transformers (CTs), voltage transformers (VTs), smart meters, humidity and temperature sensors, accelerometers, pyranometers and pyrheliometers (for solar irradiance measurement), internet protocol (IP) network cameras, power quality monitors and many more \citep{ref75}.

Control and Monitoring applications require various types of information, such that the sensor data should meet the expected distribution network operations. Some requirements of SG sensors are:

\begin{itemize}
    \item Accurate synchronization of sensors to Coordinated Universal Time (UTC) \citep{ref20}.
    \item Exceedingly accurate sensitivity and accuracy of measurements, such as for voltage/current magnitudes and phase angle \citep{ref20,ref77}
    \item Rapid processing of intelligent algorithms and data, such as producing synchronized frequencies, phasors, and rate of change of frequency (ROCOF) estimates \citep{ref20,ref76}
    \item Fast, reliable, and secure standards-based data transmission and network communications \citep{ref20}.
    \item Sensor variety featuring dynamic range and high bandwidth, such as measuring frequencies, voltages, and currents accurately across a wide range \citep{ref20,ref78}
    \item Smart capabilities, including self-localization, self-
    identification, self-calibration, self-diagnostics, and \newline
    self-awareness \citep{ref20,ref78} 
    \item A multitude of sensing capabilities for physical (temperature, climate, weather, etc.) and electrical (power flow, current, voltage, etc.) parameters \citep{ref20}.
    \item Standardized testing methodologies and interfaces to assist in achieving Smart Sensor (SS) interoperability and plug-and-play capabilities \citep{ref20}.
\end{itemize}

\subsection{Communication and Data Exchange}

\begin{figure}[ht]
\centering
\includegraphics[width=8.5cm, height=6cm]{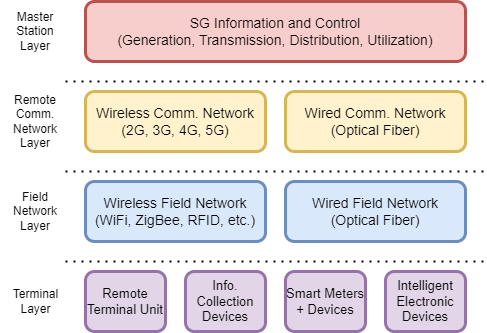}
\caption{SG Communication Layers \citep{ref88}}
\label{fig:com-layer}
\end{figure}

Effective communication is critical to the successful deployment of SG. The domains and sub-domains of the substations will use an assortment of private and public, wired, and wireless communication networks to interchange information \citep{ref79}. In general, telecommunication infrastructure is crucial, however wireless communications offers much more freedom for information collection, dissemination, and processing. Many suggestions for communication methods have been made and some of the most commonly considered communication technologies are ZigBee, Wireless Mesh, and Power Line Communication (PLC). ZigBee is a short-range wireless communication used in home area networks \citep{ref80}, wireless mesh network communication connects multiple devices as nodes in a larger network \citep{ref81}, and PLC is a wire line communication technique that is applied at very low costs \citep{ref72}. Some relatively new, less explored communication options are detailed as follows. Figure \ref{fig:com-layer} below also provides a representation of the multiple communication layers involved in SG.

\subsubsection{Wireless Sensor Network (WSN)}

Wireless sensor networks (WSNs) are able to construct an extremely reliable power grid with self-healing capabilities, necessary in the deployment of SGs \citep{ref74}. Modern advancements in WSNs have enabled the development of embedded electric utility monitoring systems \citep{ref82}. Some potential applications in the SG include fault sensing, remote monitoring, and wireless automatic meter reading (WAMR) \citep{ref82,ref83}. WSNs can be implemented and utilized across the whole SG network because of its flexibility, rapid development, and low cost \citep{ref74}.

\subsubsection{SG Interoperability Platform (SGIP)}

To address common interoperability issues, the novel SGIP communication method consists of managing data-driven and communication-driven interoperability, enabling data-centric communication in SG \citep{ref84}. The data-driven interoperability is guaranteed through the implementation of a common semantic model, ensuring systems communicate in the same way (or, "in the same language") \citep{ref84}. The SGIP design is based on many questions and several requirements, including support of both publish-subscribe and client-server communication, support of reconstructing data objects, and configurable quality of service (QoS) attributes for reliable communication \citep{ref84}. Efforts have been made by NIST as well to establish a SG interoperability panel (also called SGIP) and interoperability in smart sensors \citep{refba}.

\subsection{Communication Solutions}

Table 5 presented below, summarizes the papers related to communications as examined in this section.

\begin{table}[ht]
\centering
\caption{Summary of Communication Solution Papers}
\label{tab:my-table3}
\resizebox{\columnwidth}{!}{%
\begin{tabular}{l|l|l}
Year & Reference & Method  \\ \hline
\rule{0pt}{4ex}
2012 & \citep{ref85} & \begin{tabular}[c]{@{}l@{}}Wireless communications for advanced metering\\ infrastructure.\end{tabular} \\
\rule{0pt}{3ex}
2013 & \citep{ref86} & Cloud computing model \\
\rule{0pt}{6ex}
2011 & \citep{ref87} & \begin{tabular}[c]{@{}l@{}}SG communications architecture three different \\ operating modes: relay control, distribution, \\ and home levels.\end{tabular} \\
\rule{0pt}{5ex}
2020 & \citep{ref88} & \begin{tabular}[c]{@{}l@{}}Joint 5G-IoT frameworks in SG and the \\ associated enhancement in interoperability.\end{tabular} \\
\rule{0pt}{4ex}
2021 & \citep{ref89} & Distributed Data Interoperability Layer (DDIL) \\
\rule{0pt}{4ex}
2018 & \citep{ref90} & Data distribution system-based approach. \\
\rule{0pt}{6ex}
2021 & \citep{ref91} & \begin{tabular}[c]{@{}l@{}}Merging of two common communication \\ systems in both IoT and SG \\ domains: oneM2M and IEC 61850 \end{tabular}  \\
\rule{0pt}{6ex}
2020 & \citep{ref92} & \begin{tabular}[c]{@{}l@{}}An approach to drive semantic interoperability, \\ led by NIST, the DOE, and multiple other \\ national US laboratories.\end{tabular}
\end{tabular}%
}
\end{table}

The authors in \citep{ref85} proposed the application of wireless communications for AMI. Focused on efficiency and security, they propose a method that enables smart meter transmission when the large changes are exhibited, referred to as Change and Transmit (CAT) \citep{ref85}. It was based on data that implies that the real-time power consumption of households is most often constant and the change of power use follows the Poisson distribution \citep{ref85}. From this trend, however, an attacker could eavesdrop and know when a user is home. Considering this, the CAT method also utilizes an Artificial Spoofing Packet (ASP) to send false packets to attackers \citep{ref85}. The scheme proposed by the authors accounts for the Poisson power consumption nature to administer their wireless communication infrastructure and was effective for mitigating attacks, comparing the effectiveness of adjusted defense schemes.

Markovic \textit{et al.} \citep{ref86} provides an in-depth discussion on a cloud computing model for SGs, where the delivery of computing is introduced as a service. This means that information, software, and storage are provided to devices as a product. Cloud computing is an architecture deemed suitable for SG applications, since its services can achieve storage and transfer of data, communication, and real time computation at substantial scales \citep{ref86}. The authors highlight the benefits and disadvantages of cloud-based schemes, where for instance, its dependence on internet connectivity poses operational and security issues \citep{ref86}. Additionally, specific applications and use-cases are outlined in the scope of cloud computing in SG, with an emphasis on modern systems (state of the art a decade ago) and future challenges that have been the subject of research since. 

Proposed in \citep{ref87} is an SG communications architecture using modern wireless communication technologies, operating in three different modes: relay control, distribution, and home levels. The distribution mode is cooperative, and because of its structure, it can dynamically solve complex coverage problems with low losses and costs, and unified resources, suitable for SGs \citep{ref87}. The relay-based system is also cooperative, featuring a node structure that can communicate with other nodes directly or by picking off its connection to another node \citep{ref87}. This can also be thought of as the wireless mesh discussed in \citep{ref81}. The home or indoor mode is short range, but can achieve high quality and fast connections using little power, similar to a WiFi network \citep{ref87}. Exhibiting low latency and high bandwidth characteristics, the architecture proposed that encompasses all three nodes was effective in several video monitoring and on-demand experiments, scalable for SGs. 

Researchers in \citep{ref88} presented an introduction to 5G communication networks and the Role of Joint 5G-IoT Framework for SG Interoperability Enhancement. They analyzed the fundamentals of 5G communication and the 5G-IoT purposes in SG. From their review, it was concluded by the authors that the development of communication infrastructure with 5G-IoT has positive impacts on the flexibility, autonomy, reliability, speed, security, and economy of operation of the SG \citep{ref88}. Speed in SG communication is a crucial benefit, where cybersecurity and stability are improved with decreased transmission delays.

In their research, A. Alaerjan \citep{ref89} presented a Distributed Data Interoperability Layer (DDIL). This layer is a connectivity layer, enabling seamless data exchange between applications in IoT domains. Using model transformation techniques, the DDIL addresses data interoperability issues. It is "developed into a set of configurable features to support the flexibility requirements in IoT applications" \citep{ref89}. The DDIL was designed to be modular and extendable, and develop four applications based on different protocols to test its effectiveness in a simulated SG environment \citep{ref89}. Experimental results demonstrated that the proposed DDIL successfully allows seamless data exchange between the communicating applications, even on constrained devices \citep{ref89}. A. Alaerjan \textit{et al.} \citep{ref90} also detailed a data distribution system-based approach for addressing interoperability challenges that exist in SG. A unified data model was adopted to build data distribution system topics, acting as the primary data exchange medium. Their experiment corroborated the viability of the data distribution system and its potential in supporting the interoperability of protocols \citep{ref90}. 

Researchers in \citep{ref91} proposed a novel solution towards the unification of two of the most commonly applied communication systems in IoT and SG domains: oneM2M and IEC 61850. Their semantic interoperability solution is primarily based on the concept of common ontology between different ontologies and oneM2M ontology-based interworking \citep{ref91}. It was also proposed to be applied to the interworking strategy. A detailed illustration of the architecture was provided, consisting of the IEC and IoT servers, the IoT domain, the interworking proxy application entity (IPE), and the proposed ontology \citep{ref91}. Cavalieri's proposal forwards the advancement of IoT implementation in SG, contributes to the limited scope of literature focused on IPE, and expands on the oneM2M definition \citep{ref91}. 

Presented in \citep{ref92} by Bergmann \textit{et al.} was an approach to drive semantic interoperability, led by NIST, the DOE Building Technologies Office, and several other US national laboratories. Their approach consists of the following steps: a semantic interoperability standard to regulate and identify the interoperable attributes of applications and equipment, industry coordination and engagement, and tools to assist in testing and implementation, ensuring product compliancy with semantic interoperability specifications \citep{ref92}. Their approach accelerates the timeline for the adoption of other semantic interoperability specifications \citep{ref92}. 

\subsection{Standards}

Standards help to ensure interoperability in SGs. For instance, IEEE 1815-2012 (DNP3) is a set of communications protocols applied between elements for the automation of process systems \citep{ref20}. DNP3 plays a vital role in SCADA systems, developed for communications between different types of control and data acquisition equipment \citep{ref93}. There are lists of standards presented in the document created by the NIST based on comments from stakeholders, public review, and workshops \citep{ref74}. The two key guidelines used to designate standards are if they support the interoperability of SG in its evolution, and the standard's level of support. For example, IEEE 1588 is used for time synchronization and management of equipment across SGs, and IEC 61850 Suite is for communications within transmission and distribution sectors, as noted by NIST \citep{ref94}.

\subsection{Testing}

Although standards help with some interoperability challenges presented in SG, issues can still arise. Interoperability testing is key to achieving seamless interoperability of SG applications, due to the complexity involved in aspects of modern power systems \citep{ref95}. Interoperability testing is used to verify that multiple devices and/or systems are capable of interoperability based on the same standards, preventing against the failure of a system with non-standardized equipment \citep{ref96}. Table 6, seen below, summarizes the papers related to interoperability testing examined in this section.

\begin{table}[ht]
\centering
\caption{Summary of Interoperability Testing Papers}
\label{tab:my-table4}
\resizebox{\columnwidth}{!}{%
\begin{tabular}{l|l|l}
Year & Reference & Method \\ \hline
\rule{0pt}{4ex}
2018 & \citep{ref97} & \begin{tabular}[c]{@{}l@{}}Analysis of current methods and principles of \\ interoperability testing. Proposes a method for \\ developing SG interoperability tests.\end{tabular}  \\
\rule{0pt}{4ex}
2017 & \citep{ref20} & \begin{tabular}[c]{@{}l@{}}Passive interoperability test scheme for smart \\ sensors, ensuring interoperability of sensor data.\end{tabular}  \\
\rule{0pt}{8ex}
2022 & \citep{ref98} & \begin{tabular}[c]{@{}l@{}}Procedure for modeling the interoperability of \\ smart sensor interactions, while applying finite \\ state processes and labeled transition systems to \\ automatically and quantitatively measure and \\ assess the interoperability. \end{tabular} \\
\rule{0pt}{6ex}
2020 & \citep{ref95} & \begin{tabular}[c]{@{}l@{}}Testing methodology proposed in \citep{ref47} and \\ investigates a flexibility activation mechanism in a \\ power grid system.\end{tabular}              
\end{tabular}%
}
\end{table}

The authors of \citep{ref97} presented an analysis of current methods and principles of interoperability testing, and proposed a method for developing/prototyping SG interoperability tests. Papaioannou \textit{et al.} describes the method as a series of steps, the first being identifying the use case of the method, with the second and third involved in determining the BAP and BAIOP profiles \citep{ref97}. The folowing steps are involved in the experiment's statistical design, interoperability testing, and final analysis \citep{ref97}. The method provided a modern theoretical analysis of working methods proposed for addressing SG interoperability. 

Researchers in \citep{ref20,refsa} proposed a passive interoperability test scheme for smart sensors to ensure the interoperability of sensor data. Passive methods are typically generalized as interoperability fault monitoring methods, whereas active methods apply some stimuli to the system to test what normal operation is \citep{ref95}. The research describes the capabilities of smart sensors, as well as the model's compatibility with PMU \citep{ref20}. The method was tested, and the results demonstrate the functionality of the proposed interoperability test scheme. Passive methods are also explored in \citep{refta}.

The authors of \citep{ref95} used the testing methodology proposed in \citep{ref97} and applied it to a specialized SG use case. The situation examines a flexibility activation mechanism and flexibility source interactions in power grid systems (including DSO SCADA) and Remote Terminal Units to support a voltage regulation service. The method incorporates a design of the experiment and testing procedures described by Papaioannou \textit{et al.} \citep{ref97}, using a physical test bed that simulates the power grid and communication network \citep{ref95}. Experimental results proved the applicability of the procedure from \citep{ref97} for testing the large scale interoperability of more complicated SGs. It was determined that rigorous analyses of statistics are required for a proper investigation of interoperability performance and parameter interactions \citep{ref95}.

Song \textit{et al.} \citep{ref98} describes a procedure for modeling the smart sensor interoperability considering interactions. Applied to a case study of the interaction between an IEEE C37.118 PMU-based sensor and phasor data concentrators, the model was designed generically as a message operation between sender/receiver \citep{ref98}. They used finite state processes and labeled transition systems to quantitatively and automatically assess and measure interoperability \citep{ref98}. The model worked to improve interoperability by identifying and resolving interoperability issues in the case study, without a time constraint \citep{ref98}. Based on other standard protocols, their model could additionally be applied for interoperability modeling of other smart sensors. 

\section{Renewable Energy Integration (REI)}

While the traditional power grid is comparatively limited, the electrical energy flow and generation in SG is considerably more adaptable \citep{6099519}. SG allows for safe integration of renewable energy resources into the grid, supplementing the power supply with the power generated and stored by a consumer \citep{ref17}. Modern grids are also capable of this net metering system in some places, such as for homes that have solar panels. The utilization of renewable energy sources in the electrical grid is rapidly increasing, due to a need for reduced dependency on conventional energy sources (i.e., fossil fuels) and high power demands. The integration of renewable energy sources into SG is being continuously studied and advanced due to its realized importance \citep{ref99}.

\subsection{Challenges}

Renewable energy source integration into the power grid poses multiple technical issues \citep{ref100}. These include reliability, efficiency, energy conversion cost, power quality, security, safety, and the appropriate management of loads \citep{ref101,ref102,ref111}. Renewable energy generation is incredibly variable, and the quantity of power generated is greatly impacted by the weather forecast. Integrating variable generation presents many unique challenges to the performance of the power grid, including key factors such as the power movement type and generator design, expected run types, the position of the grid in relation to renewable energy sources, the interactions between different renewable energy sources, and the general characteristics of the grid \citep{ref101,ref111}. There are many significant challenges SG system developers \citep{ref103}, with plenty of research being conducted about current and future issues. After surveying current and future applications for SG renewable energy integration, \citep{ref104} ensured that electronic device communication is a crucial technology to integrate renewable energy sources.

\subsection{Benefits}

Despite there being numerous challenges, there are also multiple positive impacts that the integration of renewable energy will have on the grid system. First, is the positive environmental impact. Renewable energy integration within the grid enables less dependency on energy generated by fossil fuels which will reduce carbon dioxide emissions \citep{ref105}. The impact on the climate can be minimized by making it simpler to incorporate renewable energy sources, where SGs leverage automation systems, usage data and models, and bidirectional communications for improved efficiency \citep{ref17}. Second, there is a positive social aspect. When people utilize their own energy production systems (such as rooftop solar panels), they can receive rebates for generating their own energy, as well as having a contingency if there is a power grid failure \citep{ref105}. Finally, there is a positive economic aspect as increased integration of renewable energy will create new jobs.

\subsection{Variability}

Non-renewable energy sources are typically controllable and reliable and generate a consistent amount of energy \citep{ref118}. Unfortunately, renewable energy sources do not generate a stable quantity like non-renewable sources and are heavily influenced by a variety of conditions. Consequently, one of the main challenges of using renewable energy sources are their intermittency and stochastic behaviour \citep{ref106}. Renewable energy sources are fundamentally different than non-renewable generation since the production is uncertain, intermittent, and not dispatched (unable to be controlled on demand) \citep{ref107}. Variability is the term used to encompass these three characteristics and presents one of the most important obstacles of the deep penetration of integrating renewable generation into the power grid system \citep{ref108}. Penetration in the context of renewable energy and SG refers to the percentage of electricity generated for a SG by a renewable resource, where in deep penetration, a high percentage of total energy is generated by these sources. Deep penetration is a more recent vision for the SG, but the limited predictability and variability of these renewable energy sources have presented a myriad of technical challenges for grids \citep{ref109}. These challenges include methods of energy storage management, effective forecasting, power system stability, voltage control, and demand management systems \citep{ref110,ref111,ref112,ref113}, \citep{ref24}. 

Modern operations of the electrical power system are designed to accommodate the natural load demand variability, as well as adapt to both unplanned and planned contingencies, at different timescales \citep{ref107}. This is performed through implementing operational reserves, load frequency control, unit and scheduling commitment, load shedding, and demand response \citep{ref107}. At deep penetration levels, renewable energy sources add significant variability, not capable of being handled by the current system. Previous research has suggested that the existing management mechanisms could only handle renewable generation up to 20\% penetration levels \citep{ref114,ref115}. The reconfiguration of these mechanisms comes with significant operational and cost issues \citep{ref116}. 

Solar and wind energy pose the main problems for deep penetration of renewable energy integration, since hydroelectric, geothermal, and biomass energy sources are considered, by their nature, relatively more predictable and stable, having no notable issues with SG integration \citep{ref101,ref111}. This is mainly due to the fact that they are mature processes, but more importantly, power companies have near-full control over the fuel source.

\subsubsection{Solar}

Solar power generation has a natural daily cycle of variability, and beyond this, weather changes such as cloud cover can significantly affect the source's power output. For instance, solar insolation can vary by more than 80\% of its peak in a very short time \citep{ref117}. Considering this, the absorption of solar rays are needed for solar panels to generate electricity, hence it is essential to optimize both the size and direction of the panels to maximize energy generation \citep{ref118}. Additionally, it is vital to consider other factors such as the maximum power output of the panel and supporting devices such as inverters. Also, solar panel performance depends on several factors such as the quantity of sunlight, air density, temperature, and efficiency \citep{ref118}.

\subsubsection{Wind}

Wind is another renewable energy source with high variability. Wind speed and direction are consistently unpredictable due to weather changes and seasons. Severe weather can result in wind turbines operating at unsafe speeds, forcing a shutdown, and it is also difficult to obtain accurate and reliable forecasting of wind power production \citep{ref107}. To achieve the integration of wind power in SG, four conditions must be met: the phase sequence of the power frequency and the wind power frequency must be close to the grid, the terminal voltage magnitude must match the grid, and the phase angle between the two voltages must be within five percent \citep{ref119}.

\subsubsection{Forecasting}

Challenges in grid operations with deep penetration of renewable energy sources become far more manageable if accurate solar and wind energy production forecasts are available beforehand \citep{ref107}. Meteorology plays a substantial role in improving renewable energy integration into the grid network. Knowing the relevant long-term weather patterns is needed to develop a smarter power grid \citep{ref111}. Accurate forecasting can mitigate the negative effects of integrating renewable energy sources and can allow for statistical correlations between meteorological hazards and production \citep{ref120}. In addition to scheduling systems, forecasting accuracy is essential for establishing sustainable load management systems, and for using renewable resources appropriately \citep{ref111}. 

Accurate weather forecasting can alleviate some of the variability issues encountered by wind and solar. As the accuracy of the weather forecast increases, the prediction of how much energy will be generated by renewable sources will become more accurate, reducing uncertainties in generation. Weather forecasting has existed for centuries, however, more modern methods are being developed to produce more accurate forecasts. Hewage \textit{et al.} \citep{ref121} proposed an effective deep learning-based, fine-grained weather forecasting model. For forecasting, the proposed model utilized multiple layers that used surface weather parameters across a time span. Their experiment showed the model produced better results than the typical weather research and forecasting (WRF) model, proving its accurate forecasting potential up to 12 hours \citep{ref121}. Also for SG, J. Wang \textit{et al.} \citep{refca} developed an accurate deep learning model for solving the difficult problem of wind speed forecasting. They demonstrated higher precision predictions than other models using their two-stage data processing method \citep{refca}. 

\subsection{Power Quality}

There are many potential technical challenges faced with increasing the penetration of renewable energy sources that impact power quality. These challenges include voltage fluctuation, reactive power, power system transients and harmonics, switching actions, electromagnetic interference, synchronization, low power factor, long transmission lines, storage systems, forecasting and scheduling, and load management \citep{ref101,ref102,ref122}. Any device connected to the electrical grid must meet the required power quality standards \citep{ref111}. 

\subsubsection{Voltage Fluctuation}

Voltage fluctuation is a substantial issue of solar- and wind-based energy because of the intermittency of generation \citep{ref111}, reducing the longevity of most equipment since the fluctuation disturbs sensitive electronic components \citep{ref123,ref124}. Periodic disturbances to network voltages, known as flicker, are a significant issue seen in weak grids. The degree of flicker is quantified by the allowable difference in voltage in terms of frequency and the short-term severity level \citep{ref111,ref125}.

\subsubsection{Harmonic Distortion}

Non-linear appliances with power electronics inject harmonics into the grid, which have the potential to induce voltage distortion issues \citep{ref111}. To keep the total harmonic distortions to an acceptable level, these operation harmonics must be minimized. This possible by using more robust control algorithms in current control loops \citep{ref101,ref111,ref125}. According to IEEE standards, power system harmonics should be limited in two ways: by the harmonic voltage supplied by the utility to any customer at the point of common coupling (PCC) and the harmonic current injectable by users into the utility system at the PCC \citep{ref111,ref126}.

\subsubsection{Devices}

Various devices and control systems can be implemented into the grid to help mitigate the power quality challenges. In order to minimize potential challenges, efficient power converter design, which considers the physical and dynamic properties of individual energy sources, is essential. Inverters are designed for the purpose of integrating distributed generation into the grid, while maintaining power quality standards \citep{ref126}. If they are, however, not applied correctly, additional harmonics could be introduced through high frequency switching of the inverters. By designing electronics to feature control systems, power quality can be improved while mitigating voltage fluctuations and harmonic distortion \citep{ref126}. Custom devices such as series active power filters, shunt active power filters, and a hybrid of the two are some of the latest developments, overcoming both voltage and current disturbances by absorbing the load or by compensating the harmonic and reactive power generated \citep{ref102,ref123,ref124,ref126}.

\subsubsection{Synchronization}

The synchronization of grid phase, frequency, and voltage is a popular area of research towards power quality control. One of the most common methods is based on phase-locked loop, a control method where input and output signals have related phases \citep{ref111}. Other techniques for synchronization, described in \citep{ref102}, include using multiple filters paired with a nonlinear transformation or detecting the zero crossing of the grid. However, these methods are not as popular or well performing as phase-locked loop control, where filtering introduces feedback delays and zero crossing methods could be incorrectly triggered under the presence of noise \citep{ref102}.

There are multiple proposed solutions to help improve power quality in the SG. Gandoman \textit{et al.} \citep{ref127} reviews the use of Flexible AC Transmission Systems (FACTSs) to improve multiple aspects, including power quality, of the SG. They examined that distributed FACTSs help to improve not only power quality, but also energy utilization, power factor, and ensuring energy efficiency. Ceaki \textit{et al.} \citep{ref128} examines the issues of harmonic disturbances caused by solar power and electric vehicles on the grid and proposes a solution of connecting a passive filter in each of the two systems. They concluded that the harmonic spectrum decreased, and the voltage waveforms improved after including the passive filter. 

\subsection{Demand Management}

Provided the intrinsic irregular nature of renewable energy sources, maintaining an appropriate power supply requires a load demand management system \citep{ref111}. This increases the overall efficiency and quality of the system, and can be achieved by scheduling, shifting, and/or reducing demand \citep{ref101,ref111}. In reducing demand, peak demand can be reduced, resulting in a smoother demand profile, as well as reduced costs and improved reliability.

Many of the methods of improving demand management focus on forecasting demand and incentivizing consumers to use during off-peak times. Gope and Sikdar \citep{ref129} proposes a scheme for forecasting power demand in SGs based on masking-based spatial data aggregation. The authors concluded that their scheme can ensure improved computational efficiency and privacy protection in comparison to existing solutions. Making use of deep Q-learning, Razzak \textit{et al.} also develop a prediction model for consumer electricity prices and demand, outperforming current models through iterative and data-based learning \citep{refpa}. Further load forecasting machine learning methods are also presented in \citep{refqa,refra}. Kishore and Snyder \citep{ref130} proposed an optimization model for determining the amount of time an appliance is used for to take advantage of the lower rates occurring on off-peak periods. An enhanced energy management controller (EMC) optimization model that accounts for the capacity constraints of potential electricity was also introduced. Bani Ahmad \textit{et al.} proposed an alternative demand management optimization system based on machine learning, able to maintain efficiency in energy consumption and resiliency against malicious agents \citep{refoa}. The forecasting models based on data generally are formulated using the training diagram presented below in Figure \ref{fig:train}.

\begin{figure}[ht]
\centering
\includegraphics[width=5.5cm, height=6.5cm]{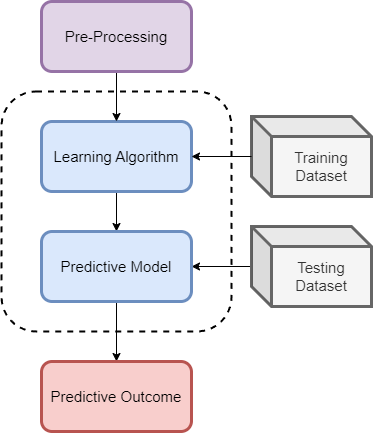}
\caption{Forecasting Model Training Diagram \citep{refoa}}
\label{fig:train}
\end{figure}

\subsection{Energy Storage}

Storage systems are designed to deliver short-term power which is used for frequency and voltage support, and renewable generation variability and power quality correction \citep{ref131}. Energy storage systems are a fundamental part of incorporating renewable energy resources into SGs. The variability of such resources can be reduced since these systems store excess energy during off-peak periods and supply it during peak hours \citep{ref15}. 

Even though there are a large number of technologies available for storing energy in SGs, many are neither economical or efficient. Efficient storage technologies are necessary for the reliability of electric power systems \citep{ref132}. Researchers analyze the utility of energy storage with respect to peak shaving, frequency stability, voltage support, transmission upgrade deferral, and renewable firming \citep{ref132}. 

Energy storage systems provide a method of delivering and storing energy to for when renewable energy sources are not able to be generated and ensuring a balance between supply and demand \citep{ref133}. They are also important in avoiding the wasting of the generated renewable energy \citep{ref118}. Storage systems are able to supply power until a low state of charge (SoC) is reached, where the primary energy source then activates to charge the storage system \citep{ref133}. There are a variety of storage technologies which need to be chosen accordingly for particular applications \citep{ref134}. 

Considering the perspective of renewable energy integration, the main consideration is determining when energy should be co-located at generation sites or distributed across the grid \citep{ref107}. Co-located storage devices allow for the possibility of real-time control, reducing the variability of the renewable energy output. It is therefore necessary to determine the minimum storage size that is needed to attain a certain extent of probability for satisfying the desired output power profile \citep{ref107}.

Tan \textit{et al.} \citep{ref135} extensively examines and reviews various energy storage technologies and their applications with renewable energy integration, including electrochemical, mechanical, thermal, and electromagnetic systems. They examine the advantages, disadvantages, and applications of energy storage technologies, commenting on the future research and goals of the most promising considering renewable sources. Pang \textit{et al.} \citep{ref136} discusses the advantages of using plug-in hybrid or battery-based electric vehicles as a means of energy storage for outage and demand side management. The energy storage method is dynamically configurable and dispersed, operating as a vehicle-to-building system \citep{ref136}.

\subsubsection{Flywheel Based Energy Storage Systems}

Stored as kinetic energy, flywheel technology efficiency is approximately 99\% effective and ideal for large-scale regulation purposes, with standby losses ranging from only 0.2 to 2\% \citep{ref13,ref138}. It has a rapid dynamic response, long life expectancy, and relatively high characteristics of self-discharge rate, energy density, power, cycling rate, and energy conversion efficiency \citep{ref132}. For short periods of time, the flywheel can administer peak power and fast responses used for balancing grid voltage sag correction and frequency oscillations \citep{ref137}. Additionally, during grid disturbances, they provide highly reliable ride through for critical loads \citep{ref138}. Some limitations remain however, where since they can only have these benefits on loads for short periods of time, they must be used in conjunction with other storage devices \citep{ref138}. Additionally, they are difficult to install, and operation standards are limited \citep{ref138}. Arghandeh \textit{et al.} simulate flywheel storage methods for a facility microgrid, demonstrating the aforementioned benefits \citep{ref138}.

\subsubsection{Hydrogen Based Energy Storage Systems}

In this system, the electricity is generated through reverse electro-chemical reactions within fuel cells \citep{ref132}. Though having a low efficiency (around 50\%), these types of storage systems have adequate dynamic response, can be used for long periods of time, and have no emissions since their only by-product is water \citep{ref132}. Research conducted by Chamandoust \textit{et al.} are one such example of hydrogen storage implementation, proposing an optimization model with multiple objectives for SGs \citep{refna}. Using four case studies, results demonstrate the storage system's effectiveness in terms of operation cost and reliability, but energy consumption deviates from the desired value due to the charging mode \citep{refna}.

\subsubsection{Thermal and Electrical Energy Storage Systems}

Thermal energy is converted from electrical energy using their generation systems that generally consist of the storage heat exchanger, heating and cooling setups, and an air handling unit \citep{ref132}, but are generally categorized in terms of sensible, latent, and thermochemical storages \citep{refma}. The charging that is required for this type of system can be done centrally or locally, with load shifting also available \citep{ref139}. These systems typically feature frequency regulation, rapid response rate, high efficiency (close to 100\%), and large storage capacities. Methods and case studies in this storage type are explored in \citep{refma}.

Not widely adopted, magnetic coils, ultracapacitors, and super conductors are costly alternatives \citep{ref140}. Considering electrical systems, ultracapacitors can be used to improve their reliability and performance, resulting in high power densities, and discharging/charging capacities \citep{ref141}. They also provide additional power to the fuel cell plant during transient or peak periods \citep{ref143}. However, they cannot alone store significant amounts of energy \citep{ref143}, which is why they are typically used in hybrid structures, demonstrated in the following subsection. Shi and Crow \citep{ref142} outline mathematical, electric circuit, and non-electric circuit models, and compare their forms for equivalency.  

\subsubsection{Hybrids}

Recently, researchers have begun proposing hybrid approaches, using multiple energy storage technologies together. A hybrid ultracapacitor and battery storage system was proposed by Kim \textit{et al.} \citep{ref144} in order to provide large scale regulation services. These services are vital in maintaining grid stability, correcting for discrepancies between power supply and demand \citep{ref144}. Their method intends to lessen the use of batteries while increasing the profitability of regulation services \citep{ref144}. A model and optimization framework for this system was derived, concluding that their system could result in profit improvements at an order range of 1.16-5.44 \citep{ref144}.  

Akram and Khalid \citep{ref145} proposed a coordinated control scheme for operating hybrid systems composed of both batteries and ultracapacitors. This approach was proposed since replacing conventional generators with renewable energy generation sources could jeopardize grid stability \citep{ref145}. Considering investment, operation/maintenance, and replacement costs, the authors concluded that their framework for the hybrid system ensures expected regulation of frequencies without information losses, again resulting in increased processes to regulation service providers \citep{ref145}. 

\subsection{Energy Efficiency}

Minimizing energy losses in SG is an exceptionally important objective. Efficient power systems optimize transmission and distribution systems, and effectively manage consumption. Table 7, seen below, summarizes the papers examined in this section.

Aquino-Lugo and Overbye \citep{ref146} implemented decentralized control algorithms with agent-based technologies, aiming to minimize power losses across distribution grids. They presented two case studies for their optimal power flow (OPF) algorithm, analyzing their performance for distributed systems \citep{ref146}. Their work concluded the validity of this control approach, but only with the presence of intelligent communication methods \citep{ref146}.  

Research from Ochoa and Harrison \citep{ref100} used a multi-period alternative current OPF for determining the optimal accommodation of renewable distributed generation (DG). Their method was aimed at minimizing energy losses, also investigating trade-offs between more generation capacity and energy losses. Their method was validated in terms of optimality and loss reduction, citing simple implementation in the majority of existing distribution networks \citep{ref100}. 

Y. M. Atwa \textit{et al.} \citep{ref147} proposed a method for allocating different categories of renewable DG units optimally, minimizing the annual energy loss. This was applied to a common rural distribution system, including constraints applied to the maximum penetration limit, the feeders' capacity, voltage limits, and the discrete size of the available DG units \citep{ref147}. The authors concluded that for all scenarios and renewable resource combinations studied, annual energy losses saw significant reductions without violating constraints \citep{ref147}.  

\begin{table}[ht]
\centering
\caption{Summary of REI Challenges Solution Papers}
\label{tab:my-table5}
\resizebox{\columnwidth}{!}{%
\begin{tabular}{l|l|l|l}
Year & Reference & \begin{tabular}[c]{@{}l@{}}Challenge \\ Type\end{tabular} & Method \\ \hline
\rule{0pt}{4ex}
2020 & \citep{ref121} & \begin{tabular}[c]{@{}l@{}}Accurate \\ Weather \\ Forecasting\end{tabular} & \begin{tabular}[c]{@{}l@{}}A deep learning-based effective \\ fine-grained weather forecasting \\ model which utilized multiple \\ layers that used surface weather \\ parameters over a period of time.\end{tabular} \\
\rule{0pt}{7ex}
2010 & \citep{ref102} & Synchronization  & \begin{tabular}[c]{@{}l@{}}Zero crossing detection of \\ the grid or using multiple filters \\ paired with a nonlinear \\ transformation\end{tabular} \\
\rule{0pt}{5ex}
2018 & \citep{ref127} & Power Quality & \begin{tabular}[c]{@{}l@{}}Flexible AC Transmission \\ Systems (FACTSs)\end{tabular} \\
\rule{0pt}{4ex}
2017 & \citep{ref128} & \begin{tabular}[c]{@{}l@{}}Harmonic \\ disturbances \\ caused by solar \\ and electric \\vehicles\end{tabular} & \begin{tabular}[c]{@{}l@{}}Connecting a passive filter \\ in solar systems and electric \\ vehicle systems.\end{tabular} \\
\rule{0pt}{6ex}
2019 & \citep{ref129} & \begin{tabular}[c]{@{}l@{}}Forecasting \\ Power \\ Demands\end{tabular} & \begin{tabular}[c]{@{}l@{}}Private and lightweight\\ masking-based spatial \\ data aggregation scheme.\end{tabular} \\
\rule{0pt}{12ex}
2010 & \citep{ref130} & \begin{tabular}[c]{@{}l@{}}Demand \\ Management\end{tabular} & \begin{tabular}[c]{@{}l@{}}Optimization model for \\ determining the time of use of \\ appliances and introduces an \\ improved powerful energy \\ management controller (EMC) \\ optimization model that accounts \\ for electric potential capacity \\ constraints.\end{tabular} \\
\rule{0pt}{7ex}
2021 & \citep{ref135} & Energy Storage & \begin{tabular}[c]{@{}l@{}}Examines and reviews various \\ energy storage technologies and \\ their applications with renewable \\ energy integration.\end{tabular} \\
\rule{0pt}{8ex}
2012 & \citep{ref136} & Energy Storage & \begin{tabular}[c]{@{}l@{}}Using plug-in hybrid \\ and battery-based electric \\ vehicles for storing energy, and \\ strategy for adopting these uses \\ in vehicle-to-building mode.\end{tabular} \\
\rule{0pt}{6ex}
2017 & \citep{ref144} & Energy Storage & \begin{tabular}[c]{@{}l@{}}Hybrid ultracapacitor and \\ battery storage system in order \\ to provide large scale regulation.\end{tabular} \\
\rule{0pt}{6ex}
2018 & \citep{ref145} & Energy Storage & \begin{tabular}[c]{@{}l@{}}Coordinated control scheme to \\ operate a hybrid system composed \\ of both batteries and ultracapacitors.\end{tabular}  \\
\rule{0pt}{8ex}
2010 & \citep{ref146} & Energy Efficiency & \begin{tabular}[c]{@{}l@{}}Agent based technologies for \\ decentralized control algorithm \\ implementations, minimizing power \\ losses in the distribution grids.\end{tabular}  \\
\rule{0pt}{8ex}
2010 & \citep{ref147} & Energy Efficiency & \begin{tabular}[c]{@{}l@{}}Method applied to a common rural \\ distribution system for allocating \\ different types of renewable \\ DG units optimally.\end{tabular} \\
\rule{0pt}{12ex}
2011 & \citep{ref100} & Energy Efficiency & \begin{tabular}[c]{@{}l@{}}Multi-period alternative current \\ optimal power flow for determining \\ the optimal accommodation of \\ renewable distributed generation. \\ Minimized energy losses \\ and investigated trade-offs \\ between those losses and increasing \\ generation capacity.\end{tabular}
\end{tabular}%
}
\end{table}

\section{Industry Applications and Future Trends}

\subsection{University/Small Scale Implementation}

\subsubsection{Research/Academic Papers}

Many universities across the world have laboratories dedicated to researching SG, including Canadian Universities such as York University \citep{refia}, Toronto Metropolitan University \citep{refja}, and University of Waterloo \citep{refka}, and the University of Melbourne in Australia \citep{refla}. These laboratories have conducted a significant amount of research and published numerous academic journals and surveys on various aspects of SG. As the demand for a more sustainable method of delivering energy increases, more research is dedicated to SG. 

\subsubsection{Prototypes}

Alongside research and academic papers, prototypes of various aspects of SG, such as demand side management \citep{ref148} or remote monitoring \citep{ref149}, have been created to design and test components of the SG. Small scale grids are ideal for analysis and the testing of new technologies, where their results are scalable and their effects on city-wide scales can be extrapolated. More recently, larger scale, more complete prototypes have been created by a select number of universities and companies.

A significant milestone in SG prototyping is the award-winning Microgrid Energy Management System (MG-EMS) prototype, pictured in the Figure above. Located at the Laboratory for Clean Energy Research (LaCER), School of Electrical and Electric Engineering at Nanyang Technological University in Singapore, this prototype uses software applications for managing sensed data and performing generation and load management routines \citep{ref150}. By using NI LabVIEW software, CompactRIO hardware featuring FPGA technology, and NI DAQ hardware, they extend the microgrid to a SG prototype \citep{ref150}. A maximum power point tracking (MPPT) module of the photovoltaic (PV) system was created with this method, as well as a control scheme of the BESS to integrate energy management systems for buildings (HEMS, BEMS), solar PV technology, and energy storage with the existing microgrid prototype \citep{ref150}.  

Green Empowerment’s SGs for Small Grids project aims to bring intelligent, open-source technology to engineers and technicians in remote communities. They work with regional partners to build renewable energy micro-grids with remote indigenous communities in South Asia \citep{ref151}. Green Empowerment also develops and builds prototypes for lab testing with real-world appliances, combines with simulated challenges. They work closely with their partners to assess financial sustainability. In 2022 and 2023, they had the opportunity to test their prototype appliance controllers and monitoring devices in Malaysia and the Philippines \citep{ref151}. 

\subsection{Country/Continent Implementations}

Many countries around the world have also begun working on the SG, whether working on pilot projects or taking initiatives for testing and research. Governments of countries such as the United States, Australia, Britain, China, Japan, and South Korea have also considered SGs as a method of reducing carbon emissions and energy security \citep{ref152}.

NIST is also devoting research and attention to the cooperation of multiple countries for the development of international standards for the SG \citep{ref38}. Canada, Mexico, Brazil, Japan, South Korea, Australia, India, China are among the countries that have already or plan on investigating the substantial SG infrastructure \citep{ref27}, setting a baseline of energy standards that other countries will soon follow in. The countries and continents examined in the following section can be seen highlighted blue below in Figure \ref{fig:imp}.

\begin{figure}[ht]
\centering
\includegraphics[width=8.5cm, height=6cm]{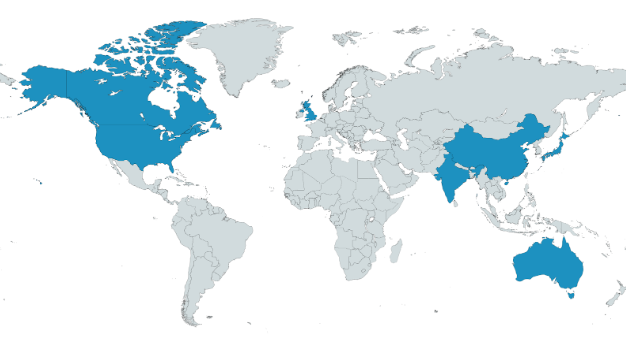}
\caption{Implementations Examined in this Paper}
\label{fig:imp}
\end{figure}

\subsubsection{Australia}

After a SG proposal in 2009, the Australian government was interested in investing \$100 million. They were also interested in raising customer awareness of energy utilization and establishing generation management and distributed demand systems \citep{ref152}. In New South Wales, five sites were chosen for SG establishments and Energy Australia working with IBM, GE Energy, and Grid Net, were selected to work on the project. The idea was to build a Worldwide Interoperability for Microwave Access (WIMAX)-based SG that had capabilities of automatic substations, able to support up to 50,000 smart meter connections and to accommodate electric vehicles \citep{ref152}. 

Most recently in 2021, the Australian Energy Security Board proposed an energy market redesign for 2025 \citep{ref153}. Built around the use of the SG and renewable energy sources, they suggested an assortment of reforms necessary for the transition to become the most decentralized power system in the world \citep{ref153}. With reforms established and enforced, the benefits will be targeted towards consumers, enabling flexibility, reliability, and affordability in Australia's power grid \citep{ref153}.

\subsubsection{India}

The Indian government has initiated nationwide SG projects, with the primary goal of identifying and discussing the deployment barriers and concerns, such as customer acceptance. Customers proved willing to adopt SG and a mini SG project was implemented by the Puducherry Electricity Department in Puducherry \citep{ref154}. This project had proposed an area covering 34,000 consumers from a wide variety of monthly incomes and electricity usage, enabling each with advanced metering technologies \citep{ref154}. Many projects have been proposed and implemented across India, in cities such as Ajmer, Tripura, Haryana, and Manesar \citep{ref154}. There are currently 12 total SG pilot projects in power distribution sectors listed, which include the adoption of various functionalities, such as advanced metering technologies, load management, substation automation including SCADA, distribution transformer monitoring, distributed generation, outage management, power quality measuring, micro grid, electric vehicle charging infrastructure, home energy management center, and cyber security and training infrastructure \citep{ref154}.

\subsubsection{United Kingdom}

The UK has actively been decreasing greenhouse gas and carbon dioxide emissions since 1990 \citep{ref155}. Currently, the UK is in the process of adopting the SG on a country-wide scale, with a phased approach led by industry regulator Ofgem and the Department of Energy and Climate Change (DECC) \citep{ref155}. In 2021 alone, 8 million smart meters were installed and connected to the Data Communications Company (DCC), who state their total 17 million units at the time help in reducing 500,000 tonnes of carbon dioxide per year \citep{ref155}. Their goal is to implement 53 million smart meters by 2025 and as of 2022, they published a report proposing reforms for the energy market, digitizing the energy system and adopting novel smart technology \citep{ref155}. Ofgem's proposed reforms aim to provide consumers increased energy consumption authority, as well as promoting net-zero green house gas emissions.

\subsubsection{China}

In September 2020, the Chinese government announced an objective of peak carbon dioxide emissions by 2030 and carbon neutrality by 2060 \citep{ref156}. Since then, they have been trying to build and improve their energy system \citep{ref156}. China is expected to lead in terms of advanced metering infrastructure deployment, as they have begun replacing their first generation of smart meters with more advanced systems \citep{ref157}. 

\subsubsection{Canada}

There are SG programs happening in several provinces and territories throughout Canada. The SG program, one of Natural Resource Canada’s programs, funds \$100M over four years (2019 to 2023) to progress the demonstration of SG technologies and the deployment of SG integrated systems \citep{ref158}. During the four-year program, recipients reported on the initial deployment of the grid and its influence during their project up to 5 years following their project investment \citep{ref158}. The information reported will be used by Natural Resource Canada to analyze grid impacts and program future lessons to inform of future programs and policy development. Currently, the SG program has funded 21 projects across Canada working on development projects, deployment projects, or hybrid projects, as seen in Figure \ref{fig:can} \citep{ref158} below.

\begin{figure}[ht]
\centering
\includegraphics[width=8.5cm, height=6cm]{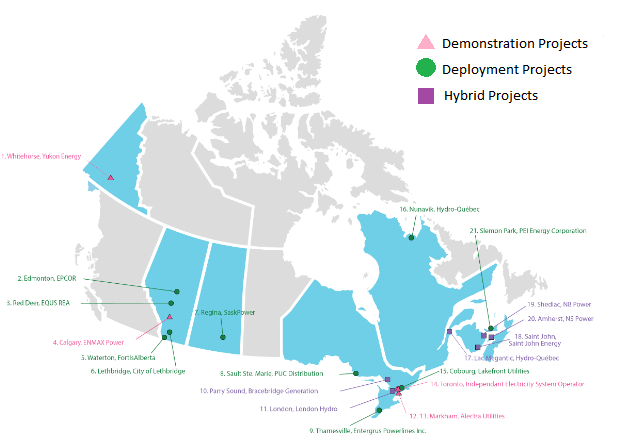}
\caption{SG projects throughout Canada}
\label{fig:can}
\end{figure}

Additionally, Canada's Energy Innovation Program and Smart Grid Demonstration provides funding to projects that demonstrate innovations in smart grid solutions or technologies \citep{refha}. Having just closed its call for proposals this year, the project promotes accelerating grid modernization, improving customer accessibility, addressing market gaps, and advancing diversity, inclusively, equity, and accessibility in this field \citep{refha}

\subsubsection{United States}

The US DOE was provided with \$4.5B from the Recovery Act for the modernization of the power grid \citep{refga,ref159}. The two largest initiatives are the SG Demonstration Program (SGDP) \citep{reffa} and the SG Investment Grant (SGIG) \citep{refga,ref159}. The SGDP works on advanced SG and energy storage systems and evaluates performance for future applications. Overall consisting of 32 projects, 16 of those are dedicated to regional demonstrations to prove validity, quantify costs, and test scalable business models \citep{reffa}. The other 16 projects is for the energy storage system development, such as the ones discussed in the previous section. The SGIG focuses on the accelerating the deployment of existing SG techniques, tools, and technologies for improving modern grid performance, comprising approximately \$8B from the Act. The program involves 99 projects, consisting of power supply companies, who upgrade their systems to test distribution and transmission systems \citep{refga}. There are also other SG programs such as ones for work training, interoperability and cybersecurity, and renewable and distributed systems \citep{ref159}. 

\subsubsection{Japan}

In 2015, the Japanese government identified six portions vital to accelerating the development and deployment of SG in the country. These key portions included smart meters as the most vital, along with, communications technologies, solar PV, battery energy storage systems, energy management systems, and the deregulation of the electricity market \citep{ref157}. The countries efforts towards SG were accelerated amid hosting the 2020 Olympic games as they aimed to ensure a secure energy supply. COVID-19 disruptions delayed the countries’ SG plans, and so their efforts are projected to continue through 2030 \citep{ref157}. A book from Ida \textit{et al.} outlines the current economics of SGs in Japan, its developments, and the results of four field experiments conducted in different regions \citep{refea}. The field experiments, performed in Kitakyushu city, Keihanna Science City, Toyota city, and Yokohama city, were designed to be "randomized control trials" on real power grids in residential areas \citep{refea}. The validity of the experiments is upheld by this random aspect, as well as findings regarding behavioural economics and the use of large data sets \citep{refea}. 

\subsection{Future Implementation Work}

\subsubsection{Future Research}

An abundance of research has been dedicated to the development of SGs, either directly or indirectly. The pursuit of obtaining more reliable and efficient power grids will continue as long as engineering in all fields does so as well. Especially in the fields of renewable resources and cyber security, new and more powerful/accurate methods are constantly being developed. For example, attack identification and defense methods, or more efficient power generation and storage methods are actively researched, applicable to many other applications outside the scope of SG. The well of literature exploring methods to further improve those methods is equally as vast, such as the designing of faster processors or optimizing power generation turbine blades. The advancement of interoperability and grid architectures is a by-product of the growth of cybersecurity and renewable energy integration, as well as society as a whole. As new technologies are developed, the problem of unifying them with the power grid or maximizing functionality will always exist. As humans observe the changes to the climate and population, the way energy is produced or allocated from location to location is another crucial consideration. 

Though existing portions of the SGs are well research and developed, the application of new research or modern problems must be considered under the same scope. Provided the current state of climate change and the digital age, areas that are likely to be explored more in the future include cyber security, specifically FDI attacks, improving accuracy of forecasting, micro-grid integration, and energy and demand management systems \citep{ref152}. It can be suggested provided current research in areas such as security and forecasting \citep{refca,refda} that the utilization of deep learning/large data sets will play a pivotal role in overcoming the challenges of the future.

\subsubsection{Constraints}

Cost is one of the foremost constraints challenging the further development and implementation of the SG \citep{ref160}. There is a significant cost associated with the distribution and transmission systems, as well as the additional technologies and equipment, such as smart meters. This will be an especially challenging constraint for developing countries looking to implement the SG. A study done in 2017 by Young explored SG potential in these countries, calculating the expected cost of SG and the ratio of the nation’s GDP to the cost of development. They compared factors of available resources, proximity to urban centers, the accessibility of education and training, political stability, and development cost with Kepner-Tregoe analyses \citep{ref160}. This study showed a large distribution of ratios, with the lowest being 2 and the highest being 47, demonstrating the financial challenges of implementing SG and the determination of five developing nations most compatible for investment in SGs \citep{ref160}.

\subsection{Future Implementation Expectations}

The eventual goal is for the SG to be implemented everywhere, making it the new method of transmitting and distributing electrical power. The SG will provide a more efficient method by enhancing resiliency, flexibility, and reliability of the power system, adapting self-healing grid infrastructure, improving power management, and better utilization of existing electricity assets \citep{ref158}. Implementing international standards will result in increased efficiency for manufacturers and will encourage supplier competition \citep{ref161}. This will lower costs, benefiting utilities and customers \citep{ref158}. The SG will also lead to more being available jobs throughout the energy sector. The SG is considerably more environmentally friendly than its alternative standardized power grid, providing a new solution to enable increased penetration of renewable energy generation, and reduce greenhouse gas emissions.

\section{Conclusion}

In this survey, the most prominent components and issues related to SGs are outlined and discussed. Some notable steps taken by various countries/continents towards SG implementation are highlighted, such as proposed future redesigns including the SG, investing millions of dollars into testing, developing and deploying SG technologies, and identifying and discussing actual SG deployment barriers and concerns. Finally comments are made on the current state of the SG. 

The significance of SGs was also featured, as it pertains to a more cooperative and sustainable method of power transmission and distribution. The main areas explored in this survey were the issues that pertained to cyber security, interoperability, and renewable energy integration. 

From the surveyed literature, it was evident that the challenges faced varied widely for the areas explored, all requiring further exploration and research. The threat that false data injection (FDI) attacks pose to the cyber security of SG is a more recent and dangerous threat, and is hence a popular research topic. Interoperability presents one of the main features of SG, allowing for the exchange of information. Ensuring effective communication of all components in SG is essential to realize this exchange. Renewable energy integration is of growing importance for SG, as the demand for more renewable and sustainable energy generation methods increases. Integrating these resources into the SG will be necessary to meet the energy demand and environmental challenges of the 21st century. 

The most promising directions for future research differs for the various components of SG. FDI attacks prove one of the most prominent challenges in cyber security. While multiple mechanisms have been proposed for FDI attack mitigation and defense, more research, testing, and simulations should be conducted to ensure the proposed schemes, when integrated into SG, will provide a method with high detection accuracy and resilience to attacks. Based on research the popularity and successes of data driven and deep learning methods for attack identification/rectification, it is reasonable to suggest that cybersecurity solutions involving such should be continued to be pursued, such as the CDS framework previously discussed.

While standards and methods of interoperability are a necessary component, it is also vital to develop improved interoperability testing. These tests will be vital to ensuring that components are capable of interoperability. 

Methods of controlling side effects of REI are vital to their integration into the grid, however emphasis should also be given to investigating energy storage methods. Energy storage is necessary to ensure load demand is met at all times, especially given the periodic and stochastic nature of renewable energy sources. New and innovative methods of storage, especially the emerging V2G (vehicle-to-grid) and V2X (vehicle-to-everything) methods given the growth in popularity of electric vehicles, should be further explored.

\bibliographystyle{elsarticle-num}
\bibliography{bib.bib}

\end{document}